# Stochastic Theory of the Size Distribution of Raindrops

Maksim Mezhericher and Howard A. Stone

Department of Mechanical and Aerospace Engineering, Princeton University
Princeton, NJ 08544, USA

**ABSTRACT**

For over a century, raindrop size distributions have been a subject of extensive scientific study, typically described by models including the Marshall–Palmer exponential equation, gamma, Weibull, lognormal and other mathematical functions. In this work, we present a theory that integrates deterministic principles from thermodynamics and fluid dynamics with stochastic elements to predict expected raindrop diameters and describe ground-level drop-size distributions. Importantly, our approach avoids assuming specific drop-size dispersion processes or relying on multi-variable empirical data fitting. We derive analytical equations for key raindrop parameters (e.g., median diameter, expected minimum and maximum diameters) and drop-size distributions as functions of rainfall intensity. Our theoretical predictions align well with extensive published experimental data, covering rainfall intensities from 0.4 to 40 mm/h across diverse global locations. Additionally, consistent with observations of super-large raindrops, our theory suggests a maximum ground-level raindrop diameter limit of around 10 mm. We establish analytical expressions for meteorological parameters like the Marshall-Palmer constant and total raindrop concentration, validated against empirical data. The theoretical approach presented here can find broad applications in climate modeling, sprays and aerosol dynamics, bubbles and particles, extraterrestrial rainfall, and paleoclimatology.

# I. INTRODUCTION

Rain is a natural precipitation phenomenon that has intrigued humans for millennia. The ancient Greeks believed rain is brought to Earth by Hyades, daughters of the Titan Atlas and the Oceanid Aethra [1], and the Romans thought that weather and rainfall are ruled by planets and stars [2]. Nevertheless, at that time, the Greek philosophers Aristotle and Anaxagoras described the origin of rain as vapor in air that rises due to heat, then condenses into small drops forming a cloud and precipitates back to Earth due to cold [3,4]. Later, in the Middle Ages, extremely heavy rainfalls resulted in devastating floods and crop destruction in Europe, which was reflected in chronicles [5–7] (see Fig. 1a) and drew the attention of medieval scientists who tried to explain rain and weather phenomena [4,8]. Since the 19$^{th}$ century geophysicists have found physical evidence of fossil imprints left by raindrops [9–18] at different times in Earth's history, and from the 20$^{th}$ century planetary scientists discovered signs of the existence of extraterrestrial rains on space objects in the solar system, e.g., Titan (methane rain [19–21]), Mars (water rain in the past [22–24]), Venus (sulfuric acid rain [25]), Jupiter (helium rain [26]), Neptune and Uranus (diamond rain [27,28]), as well as on some exoplanets (rains of iron [29,30], glass [31], rubies and sapphires [32], and water [33]).

In contemporary science, researchers have systematically studied rain and raindrops for more than one hundred fifty years [9,34,35] to understand the complex evolution of drops from clouds into precipitation, as well as the formation of different sizes of drops usually observed on the ground during a rainfall (see Fig. 1b). However, despite abundant studies, the complex evolution of drops from clouds into precipitation is still poorly understood [36,37]. For example, Langmuir [35] hypothesized that a "chain reaction" of drop growth by accretion and reduction by aerodynamic breakup in clouds is responsible for the formation of the observed raindrops. Subsequently, aerodynamic breakup of falling drops was investigated experimentally [38–41]. It was concluded that the drop-size distribution in heavy rains may have little dependence on microphysical processes in clouds that initiate the rainfall, and instead is governed by breakup events of big drops due to aerodynamic instability and inter-drop collisions [41,42]. More recently, the polydispersity of raindrops was explained by treating them as the products of aerodynamic fragmentation of individual non-interacting bigger drops, quantitatively demonstrating that the observed smaller break-up fragments match the measured raindrop size distributions [43–45]. For a free falling drop the aerodynamic fragmentation is achieved within a considerably shorter timescale than the typical collision time between the raindrops [45,46]. Analyzing the fluid dynamics of fragmentation of a single raindrop was also found useful in capturing the main features of extraterrestrial rainfalls [36,47,48]. On the other hand, a contrasting description exists, which is that collision-induced drop breakup is more important than aerodynamic drop fragmentation in determining the drop-size distribution in rains [49–53]. To comprehend the importance of raindrop fragmentation and collisions, we calculate below gravity, drag, capillary, collision and bursting time scales for raindrops of different sizes, and conclude if and when either of them plays a dominant role.

Here we develop a theory based on thermodynamics and fluid dynamics to quantify the stationary size distribution of raindrops, and validate the theory against a wide range of published experimental data. For this purpose, we follow the ideas of previous studies [36,40–43,45–47] that the representative diameter and polydispersity of raindrops observed on the ground is governed by the physical processes occurring after the initially formed raindrops leave the cloud. In fact, it appears that measurements (via planes, ground radars or satellites) of the initial drop-size distribution in or just below clouds may correlate poorly with corresponding measurements at the ground by meteorological stations [54–60]. In this work we do not presume *a priori* any type of drop size dispersion process or use empirical fitting to experimental data of the raindrop size distribution. First, we examine the characteristic time scales and perform dimensional and scale analyses of physical phenomena and processes occurring during the fall of raindrops. This enables us to establish governing dimensionless numbers and construct a diagram of expected raindrop diameters based on the principles of possible, impossible and expected drop diameters, which we described recently for aerosols and sprays [61]. Next, we employ mass, momentum, and energy balances, and incorporate a stochastic description to address the energy fluctuations for drops in rainfall. This leads to the derivation of underlying stochastic differential equation describing the distribution of raindrop masses at various conditions. We show that for a steady rain and white noise energy fluctuations, a particular solution of the central stochastic differential equation is the lognormal size-distribution of raindrops. We extensively validate our theory with published experimental data collected by different research groups around the globe, provide theoretical explanation why the largest observed raindrops had diameters less than ~10 mm, and find analytical formulae for Marshall-Palmer's constant and total number concentration of raindrops in air as functions of the surface tension and density of rainwater, density of air, gravitational acceleration and magnitude of stochastic energy fluctuations. Our study provides new valuable insights into the mechanisms underlying the variation of raindrop sizes by revealing atmospheric processes as a fundamental thermodynamic – fluid dynamic machine that shapes the distribution of raindrop diameters observed on the ground.

## II. THEORY DEVELOPMENT

### A. Expected raindrop diameters

We focus on a spatial portion of rain below a cloud enclosed by an imaginary surface, known in thermodynamics as a control volume [62], as shown in Fig. 1c. Developing macroscopic material and energy balances for the control volume [63], between the cloud and ground, we avoid the necessity to presume *a priori* any type of drop size dispersion process for the rainfall inside the control volume. Instead, we consider the balances of mass and energy crossing the boundary of the control volume. However, the solution of the underlying system of differential equations of material and energy balances requires additional information on the work performed by the forces inside the control volume. Here we utilize previously published ideas that single-drop fragmentation determines the raindrop size distribution, and for a free falling raindrop the aerodynamic fragmentation is typically achieved within a considerably shorter timescale than the typical collision time between the raindrops [45,46], as given in Table A1 (see Appendix: Materials and Methods). Therefore, in this model we do not explicitly include collision and dissipation caused by drop-air skin friction in the balance equations, and assume that the interplay of aerodynamic drag and gravity versus surface tension and viscous dissipation in rainwater is dominantly responsible for establishing the sizes of raindrops observed on the ground for steady non-evaporating rainfalls. At the same time, in the next section in our discussion of the size distribution of raindrops, we analyze the energy balance and introduce a stochastic term that implicitly accounts for various sources of random energy fluctuations. These fluctuations may include bouncing, breakup, and coalescence of colliding raindrops, self-oscillation of raindrops, air turbulence, and other minor stochastic atmospheric effects resulting in the spread of drop sizes around the most expected value.

Considering rain as an ensemble of multiple dripping streams, we introduce a characteristic time of a rainfall, $\tau_R = \dfrac{1}{\psi_R \hat{A}}$, where $\psi_R$ is number flux of raindrops falling on the ground (number of drops landing on a unit area of the ground per unit time) and $\hat{A}$ is a unit area of the ground surface. Next, we introduce rain frequency as the inverse of the rain characteristic time, $\omega_R = \dfrac{1}{\tau_R}$, which is an intensive parameter describing the rate at which raindrops land on the ground. From the energy balance between the air and raindrops in the control volume, we obtain that the rate of work done on the raindrops by the forces external to raindrop, $\dot{W}$, is equal to the sum of work rates of forces of air drag, $\dot{W}_D$, and gravity, $\dot{W}_g$, supplied to the drop (see Fig. 1c):

$$\dot{W} = \dot{W}_D + \dot{W}_g \qquad (1)$$

For a raindrop falling with terminal velocity, the forces of air drag and gravity are in equilibrium, and thus we can assume $\dot{W}_D = \dot{W}_g$. Therefore, $\dot{W} = 2\dot{W}_g$. On the other hand, the rate of work supplied to the raindrops is spent on the change of the raindrop surface area, $\dot{W}_\gamma$, and viscous dissipation, $\dot{W}_\nu$. Using Eq. (1), we find:

$$2\dot{W}_g = \dot{W}_\gamma + \dot{W}_\nu \qquad (2)$$

This equation can be rewritten in terms of specific (per unit mass) work rates as follows:

$$2\frac{\dot{w}_g}{\dot{w}_\gamma} = 1 + \frac{\dot{w}_v}{\dot{w}_\gamma}, \qquad (3)$$

where $\dot{w}_g$, $\dot{w}_\gamma$ and $\dot{w}_v$ are specific rates of work of gravity, surface tension and viscous dissipation forces, respectively. Using the scaling analysis (details are given in Appendix: Material and Methods) and examining the ratios of energy rate scales of the associated phenomena, we find that these work rates are functions of raindrop diameter, $d$, gravitational acceleration, $g$, rain frequency, $\omega_R$, and properties of rainwater including surface tension, $\gamma$, density, $\rho$, and dynamic viscosity, $\mu$: $\frac{\dot{w}_g}{\dot{w}_\gamma} \propto \frac{\rho g^{3/2} d^{3/2}}{\gamma \omega_R}$ and $\frac{\dot{w}_v}{\dot{w}_\gamma} \propto \frac{\mu^3}{\gamma \rho^2 d^3 \omega_R}$. Correspondingly, the dimensionless group $\frac{\rho g^{3/2} d^{3/2}}{\gamma \omega_R}$ represents the ratio of specific work rates of gravitational effects and air drag relative to the surface tension, and $\frac{\mu^3}{\gamma \rho^2 d^3 \omega_R}$ is ratio of specific work rates of viscous dissipation relative to the surface tension. Thus, Eq. (3) can be rewritten as the non-dimensional equation:

$$2\frac{\rho g^{3/2} d^{3/2}}{\gamma \omega_R} = k_1 + k_2 \frac{\mu^3}{\gamma \rho^2 d^3 \omega_R}, \qquad (4)$$

where $k_1$, $k_2$ are dimensionless proportionality constants with values of order of unity. Using rainwater properties $\rho = 1000$ kg/m$^3$, $\mu = 0.001$ Pa·s, $\gamma = 0.072$ N/m and $g = 9.81$ m/s$^2$, and typical values of $d$ and $\omega_R$ we conclude that the second term in Eq. (4) can be neglected because $\frac{\mu^3}{\gamma \rho^2 d^3 \omega_R} \ll 1$.

On the other hand, applying dimensional analysis for the six parameters $\rho$, $g$, $\mu$, $\gamma$, $d$, $\omega_R$ involved in the process of raindrop formation, and their underlying three physical units [L], [M], [T], we establish three dimensionless numbers $\pi_1 = \frac{\gamma \rho d}{\mu^2}$ (ratio of surface tension to viscous forces), $\pi_2 = \frac{\rho^2 g d^3}{\mu^2}$ (ratio of gravitational to viscous forces) and $\pi_3 = \frac{\mu}{\rho \omega_R d^2}$ (ratio of characteristic time of rainfall to characteristic time of viscous dissipation). It can be shown easily that the dimensionless groups in Eq. (4) are combinations of $\pi_1$, $\pi_2$ and $\pi_3$, i.e., $\frac{\rho g^{3/2} d^{3/2}}{\gamma \omega_R} = \frac{\pi_2^{3/2} \pi_3}{\pi_1} = \pi_2^{1/2} \cdot \frac{\pi_2 \pi_3}{\pi_1}$ and $\frac{\mu^3}{\gamma \rho^2 d^3 \omega_R} = \frac{\pi_3}{\pi_1} \ll 1$. Noting that $\frac{\pi_1}{\pi_2 \pi_3} = \frac{\gamma \omega_R}{g \mu}$, we obtain from Eq. (4):

$$\frac{\rho^2 g d^3}{\mu^2} = \frac{k_1^2}{4}\left(\frac{\gamma \omega_R}{g \mu}\right)^2 \qquad (5)$$

From either Eq. (4) or Eq. (5), we find that the expected raindrop diameter, $\tilde{d}$, is given by:

$$\tilde{d} = \left( k_1 \frac{\gamma \omega_R}{2\rho g^{3/2}} \right)^{2/3} \tag{6}$$

We will show below that from the solution of the differential equation representing an energy balance it follows that $k_1 = \frac{1}{e}$, with $e = 2.71828...$ is Euler's number, and the expected raindrop diameter $\tilde{d}$ determined by Eq. (6) is the mass median diameter [64] of mass-weighted size distribution of raindrops. Thus, from Eq. (5) we have:

$$\frac{\rho^2 g \tilde{d}^3}{\mu^2} = \frac{1}{4e^2} \left( \frac{\gamma \omega_R}{g\mu} \right)^2 \tag{7}$$

For a lognormal size distribution of raindrops (see the details in the next section), the count median diameter (i.e., the median diameter of a number-weighted drop-size distribution), $\tilde{d}_n$, is connected to the mass median diameter, $\tilde{d}$, via well-known Hatch-Choate relationship [64,65], $\tilde{d}_n = \tilde{d} \exp(-3\sigma_d^2)$. Therefore, from Eq. (7) we find:

$$\frac{\rho^2 g \tilde{d}_n^3}{\mu^2} = \frac{1}{4} \left( \frac{\gamma \omega_R}{g\mu} \right)^2 \exp(-9\sigma_d^2 - 2) \tag{8}$$

In meteorology, the rain flux on the ground is typically measured by the rain intensity, $I$, reflecting the precipitated volume of rainwater per unit ground area per unit time. The connection between the rain intensity and the rain frequency can be established by considering the overall mass flow rate of the rain $\dot{m}_R = \rho I A$, with $A$ denoting the ground area for the rain measurement. On the other hand, the rain mass flow rate is a sum of the flow rates of all mass groups of raindrops, $m_i$, i.e., $\dot{m}_R = \sum \dot{n}_i m_i = \dot{n}_R \bar{m}$, with $\dot{n}_i$ and $\dot{n}_R$ being the number per time respectively of the mass group $i$ and of the whole rain, and $\bar{m}$ is the mass of raindrops averaged over the number per time. Therefore, with number flux $\psi_R = \dot{n}_R / A$, we find that $I = \psi_R \bar{m} / \rho$. Invoking $\tau_R = \frac{1}{\psi_R \hat{A}}$ and $\omega_R = \frac{1}{\tau_R}$, we get $I = \frac{\omega_R \bar{m}}{\rho \hat{A}}$. Furthermore, considering $\bar{m} = \frac{\pi}{6} \rho \bar{d}_{mm}^3$, where $\bar{d}_{mm}$ is mass mean diameter of raindrops (average diameter based on size distribution of drop masses [64]), and employing the Hatch-Choate [65] formula $\bar{d}_{mm} = \tilde{d} \exp(0.5\sigma_d^2)$ for a lognormal drop-size distribution (more details on the distribution function are given below), where $\sigma_d$ is a standard deviation of the logarithm of raindrop diameters and $\tilde{d}$ is given by Eq. (6), we obtain:

$$I = \frac{\pi \gamma^2 \omega_R^3}{24 \rho^2 g^3 \hat{A}} \exp\left( \frac{3\sigma_d^2}{2} - 2 \right) \tag{9}$$

### B. Size distribution of raindrops

For raindrops, the material balances describe the change of flow rates of mass and number of drops when the drops pass through the control volume (Fig. 1c). The energy balance applied only to the raindrops, excluding the surrounding air, is given by the first law of thermodynamics to which we introduce an additional term representing stochastic energy fluctuations:

$$\dot{E} = \dot{W} + \dot{Q} + \dot{H}_i - \dot{H}_e + A_\xi \xi(t), \qquad (10)$$

where $\dot{E}$ is the time derivative of the energy of raindrops in the control volume, $\dot{W}$ is the rate of work supplied to the raindrops, $\dot{Q}$ is heat flow rate supplied to raindrops, $\dot{H}_i$ and $\dot{H}_e$ are rates of total enthalpy of raindrops at the inlet and outlet of the control volume, and $\xi(t)$ and $A_\xi$ are, respectively, the rate and the amplitude of stochastic energy fluctuations to which the raindrops are subjected. The rate of total enthalpy, $\dot{H}$, is the sum of kinetic, $\dot{m}_R e_k$, potential, $\dot{m}_R e_p$, and thermal, $\dot{m}_R h$, energies, where $\dot{m}_R$ denotes overall mass flow rate of the rain, and $e_k$, $e_p$ and $h$ are specific kinetic and potential energy and enthalpy, respectively. From here, the rate of total enthalpy difference between the inlet and outlet of the control volume is given by the expression $\dot{H}_i - \dot{H}_e = \dot{m}_R \left[ (e_{k.i} - e_{k.e}) + (e_{p.i} - e_{p.e}) + (h_i - h_e) \right]$. Analyzing this equation, the estimated representative values are $|e_{k.i} - e_{k.e}| \sim 0.05$ kJ/kg (for raindrop velocity increasing from 0 to ~10 m/s during the fall), $e_{p.i} - e_{p.e} \sim 10$ kJ/kg (for raindrop falling from ~1000 m above the ground) and $|h_i - h_e| \sim 84$ kJ/kg (for raindrop temperature increasing from 0 °C to ~20 °C during the fall). Thus, we can assume $\dot{H}_i - \dot{H}_e \approx \dot{m}_R (h_i - h_e)$. In addition, we consider the rainfall at constant temperature difference between the cloud and the ground, so that there is no time change of the temperature spatial profile, i.e., $\frac{\partial T}{\partial t} = 0$. Hence the supplied heat rate $\dot{Q}$ serves for sensible heating of the raindrops, and is converted into the temperature change that raindrops incur during their motion through the control volume, and effectively $\dot{Q} = \dot{H}_e - \dot{H}_i$. As a result, Eq. (10) reduces to $\dot{E} = \dot{W} + A_\xi \xi(t)$. Next, we note that we have assumed insignificant evaporation of raindrops and neglected atmospheric electric effects [66] by considering only mechanical and thermodynamical phenomena. To account for random micro-processes involved in drop size evolution and to extend the previously published literature [62,67,68] about the deterministic first law of thermodynamics applied to the complex phenomena in the control volume, we introduce a term $A_\xi \xi(t)$. This stochastic term accounts for many sources of random fluctuations of the energy of raindrops that may be caused by bouncing, breakup and coalescence of colliding raindrops [50,51], drop oscillations [66,69], internal circulation [70], vortex shedding [71], air flow turbulence [72], and other stochastic atmospheric effects. The rate of work supplied to raindrops, $\dot{W}$, is caused by the external forces acting on each falling drop. These forces include aerodynamic drag due to motion through the air column and the force of Earth's gravity field. Their combined work on a raindrop is done against the surface tension (energy required to deform a drop) and viscosity (internal resistance forces acting inside a drop) of raindrop's water. Together with random events accounted for by the stochastic fluctuation term, these interactions result in the evolution of the size of raindrops travelling from the clouds to the ground.

Combining the equations of material (given in Appendix: Material and Methods) and energy (Eq. (10)) balances for raindrops, and then expanding the terms in the differential equation to explicitly identify the dependence on raindrop mass, and employing scale analysis of the order-of-magnitude of the terms (see Appendix: Material and Methods), we derive a stochastic differential equation for the logarithm of the mass of a raindrop, $m$:

$$\frac{d}{dt}\left(\ln\frac{m}{\tilde{m}}\right) = -a\left(\ln\frac{m}{\tilde{m}}\right) + \sigma\xi(t), \tag{11}$$

where we show that $\tilde{m} = \dfrac{\pi\gamma^2\omega_R^2}{24e^2\rho g^3}$ is the median mass of raindrops. On the other hand, $\tilde{m} = \rho\dfrac{\pi\tilde{d}^3}{6}$, and thus $\tilde{d} = \left(\dfrac{\gamma\omega_R}{2e\rho g^{3/2}}\right)^{2/3}$. The latter expression for the drop diameter can be compared to Eq. (6) to provide the value of the proportionality constant $k_1 = \dfrac{1}{e}$, which we mentioned before. The coefficients $a$ and $\sigma$ in Eq. (11) reflect the physical parameters: a) $a = \dot{W}_\gamma/(2E) > 0$ characterizes the ratio between the rate of work done to overcome surface tension forces when changing the surface area of drops, $\dot{W}_\gamma$, and the energy of raindrops, $E$; and b) $\sigma = A_\xi/E$ is a measure of the energy content of stochastic fluctuations, $A_\xi$, relative to the energy of raindrops, $E$. Combining Eqs. (6) and (9), and using $k_1 = \dfrac{1}{e}$, we obtain a theoretical relationship between the mass median diameter of raindrops and the rain intensity:

$$\tilde{d} = \left[\frac{9\gamma^2}{\pi^2\rho^2 g^3}\exp\left(-3\sigma_d^2 - 2\right)\right]^{1/9} \phi^{2/9} \tag{12}$$

where $\phi = I\hat{A}$ is a volumetric flow rate of rain precipitating on a unit square area of a ground. Note that $\phi$ is an intensive parameter. The count median diameter, $\tilde{d}_n$, is obtained by multiplying the right hand side of Eq. (12) by the factor $\exp(-3\sigma_d^2)$ [64,65]. From Eq. (12) it follows that $\tilde{d} \propto I^{2/9}$, and this scaling is consistent with previously published studies [45,73–76], which demonstrated power-law dependences with the exponent value either equal [45] or close [73–76] to 2/9 (≈0.22).

The established differential equation (11) describes an Ornstein–Uhlenbeck stochastic process [77] with the logarithm of raindrop mass as a random variable. The general solution of Eq. (11) results in a family of functions describing the distribution of drop masses in rain at various conditions. One particular solution can be found by approximating the stochastic fluctuations $\sigma\xi(t)$ as a white noise produced by a sequence of serially uncorrelated random variables, assuming zero mean and finite variance of those stochastic fluctuations. In this particular case Eq. (11) can be transformed into a Fokker-Planck partial differential equation [78] for a probability density function of the logarithm of raindrop mass (see Appendix: Material and Methods). We introduce $q_m(d)$ as a mass-weighted probability density function of raindrop diameters to describe a mass-based likelihood of a drop diameter to be found between $d$ and $d + \Delta d$, and $\int_0^\infty q_m(d)dd = 1$. The mass median diameter is determined from the equation $\int_0^{\tilde{d}} q_m(d)dd = 0.5$. Then in the special case of a steady rainfall, the stationary solution of the Fokker-Planck equation leads to a lognormal form of the drop-size distribution (more details are given in Appendix: Material and Methods), as given by:

$$q_m(d) = \frac{1}{d\,\sigma_d\sqrt{2\pi}} \exp\left[-\frac{1}{2}\left(\frac{\ln d/\tilde{d}}{\sigma_d}\right)^2\right], \qquad (13)$$

where $\sigma_d = \sigma/(3\sqrt{2a})$ is a standard deviation of the logarithm of drop diameters we previously mentioned in Eqs. (9) and (12).

The theory for the mass-weighted size distribution of raindrops, Eqs. (10)-(13), leads to the conclusion that the number-weighted distribution of raindrop diameters, describing the probability of drops to have diameter between $d$ and $d+\Delta d$ based on their quantity, also follows a lognormal description similar to Eq. (13) [64]. However, for the number-weighted probability density function the count median diameter of raindrops, $\tilde{d}_n$, should be used in place of the mass median diameter, $\tilde{d}$, and $\tilde{d}_n = \tilde{d}\exp(-3\sigma_d^2)$ [64,65]. Introducing $q_n(d)$ as a number-weighted probability density function of raindrop diameters to describe a count-based likelihood of a drop diameter to be found between $d$ and $d+\Delta d$, and $\int_0^\infty q_n(d)\mathrm{d}d = 1$, based on Eq. (13) [64] we obtain:

$$q_n(d) = \frac{1}{d\,\sigma_d\sqrt{2\pi}} \exp\left[-\frac{1}{2}\left(\frac{\ln d/\tilde{d}_n}{\sigma_d}\right)^2\right] \qquad (14)$$

Here the count median diameter, $\tilde{d}_n$, is determined from the relationship $\int_0^{\tilde{d}_n} q_n(d)\mathrm{d}d = 0.5$.

It is worth noting that in many previous published reports the standard deviation $\sigma_d$ is considered as an empirical fitting parameter with values prescribed due to an absence of physical reasoning [64,75,79,80]. In this work, we make an attempt to understand more about $\sigma_d$. From the analysis of Eqs. (11) and (13), it follows that $\sigma_d \sim A_\xi$ and hence $\sigma_d$ is a measure of the magnitude of stochastic energy fluctuations, which disperse the drop sizes in rain around the expected raindrop diameter. The latter is determined by the Eq. (7) for the mass-weighted drop-size distribution, and for the number-weighted size distribution of raindrops is given by Eq. (8). The standard deviation $\sigma_d$ depends on the type of rain, rain intensity, pattern of air flow, and other atmospheric conditions having an effect on the stochastic fluctuations of the energy of raindrops. Here we employ so-called "sigma rules" of a normal distribution function [81], and determine that the logarithmic width of the raindrop size distribution lies in the range $4\sigma_d \leq \ln\frac{d_{\max}}{d_{\min}} \leq 8\sigma_d$ (see Appendix: Materials and Methods), and with $d_{\max}/d_{\max} = 10^{4/3}$ (details are given in the next section), it implies that the values of the standard deviation are confined by the range $\frac{1}{6}\ln 10 \leq \sigma_d \leq \frac{1}{3}\ln 10$. The geometric mean value of this range is $\sigma_d = \frac{1}{3\sqrt{2}}\ln 10 \approx 0.54$. Below we will examine the validity of this conclusion by comparing it with published experimental data for various raindrop size distributions.

## III. RESULTS AND DISCUSSION

### A. Comparison with experimental observations

Using the established governing dimensionless numbers $\frac{\gamma \omega_R}{g \mu}$ and $\frac{\rho^2 g d^3}{\mu^2}$ (see Eq.(5)), we construct a diagram of raindrop diameters as shown in Fig. 2. First, we plot the central line (I) corresponding to the median diameters $\tilde{d}_n$ of number-weighted size distribution of raindrops, as given by Eq. (8). Second, we draw isolines of constant raindrop diameters by varying the rain frequency $\omega_R$. Third, we plot isolines of constant rain intensities by varying the raindrop diameters $d$ and using Eq. (9) with the geometric mean of standard deviation $\sigma_d = \frac{1}{3\sqrt{2}} \ln 10$ (this value was mentioned above and more details are in Appendix: Materials and Methods) to connect $I$ and $\omega_R$. The triple crossing points between these two isolines with the central line provide the theoretical values of the median raindrop diameters $\tilde{d}_n$ at various intensities of a rainfall.

Next, we can determine the possible minimum, $d_{min}$, and possible maximum, $d_{max}$, drop diameters for a given rain intensity. Given the definitions in the balance equation of specific work rates Eq. (4), when $\frac{\rho g^{3/2} d^{3/2}}{\gamma \omega_R} \ll 1$, gravity and aerodynamic forces cannot overcome capillary forces and thus are unable to cause breakup into drops of smaller diameters, whereas when $\frac{\rho g^{3/2} d^{3/2}}{\gamma \omega_R} \gg 1$ the raindrops are always fragmented into smaller drops. Therefore, we assume that the interval of $0.1 \leq \frac{\rho g^{3/2} d^{3/2}}{\gamma \omega_R} \leq 10$ corresponds to the possible range of raindrop diameters (or possible width of the raindrop size distribution) precipitated at different rain frequencies. This interval corresponds to the region between the lower and upper boundaries of raindrop diameters given by $\frac{\rho^2 g d_{min}^3}{\mu^2} = \frac{1}{400} \left( \frac{\gamma \omega_R}{g \mu} \right)^2 \exp(-9\sigma_d^2 - 2)$ and $\frac{\rho^2 g d_{max}^3}{\mu^2} = 25 \left( \frac{\gamma \omega_R}{g \mu} \right)^2 \exp(-9\sigma_d^2 - 2)$, and indicated as lines (II) and (III) in Fig. 2. These considerations provide simple relationships between possible maximum, possible minimum and expected (median) drop diameter for a rain: $d_{min} = \tilde{d}_n \cdot 10^{-2/3} \approx 0.22 \tilde{d}_n$, $d_{max} = \tilde{d}_n \cdot 10^{2/3} \approx 4.64 \tilde{d}_n$ and $d_{max}/d_{max} = 10^{4/3} \approx 21.54$. The raindrops expected not to have diameters below the line (II) of possible minimum and above the line (III) of possible maximum diameters. Below the line (II), there is insufficient amount of available energy that can be supplied by the aerodynamic drag and gravity forces to compete with the rainwater surface tension and viscosity forces so that raindrops smaller than $d_{min}$ cannot be formed. On the other hand, above the line (III) there is substantial excess of available energy from aerodynamic drag and gravity forces over the surface tension and viscosity of rainwater, so that every drop larger than $d_{max}$ breaks apart into smaller drops. It is worth noting that the constructed diagram is based on the same principles of possible, impossible and expected drop diameters which we described recently for similar diagrams of droplet diameters of aerosols and sprays [61].

To validate the theoretical diagram of expected raindrop diameters, we plot experimental data points in Fig. 2. We utilized data reported in the literature for the average number-weighted drop-size spectrograms of more than two dozens of light (intensity $I$ <2.5 mm/h), moderate (2.5 mm/h < $I$ < 7.6 mm/h), and heavy ($I$ >7.6 mm/h) convective and stratiform rains that precipitated at different geographic locations around the world, with overall range of rain intensities 0.4-40 mm/h. The number-weighted rain spectrograms obtained by disdrometers, which are devices that measure parameters of hydrometeors on the ground, were converted into number-weighted and mass-weighted probability density functions (see Appendix and Supplementary Information). The experimental count median diameters and rain intensities were used to calculate the dimensionless groups $\frac{\gamma \omega_R}{g\mu}$ and $\frac{\rho^2 g d^3}{\mu^2}$, and plot the experimental points as shown in Fig. 2. The error bars represent measurement uncertainties that we evaluated based on the disdrometer model and measurement error analysis [82,83]. The detailed experimental reports we processed are available in the cited literature [80,84,85], and additional data references we used are given in the Supplementary Information. As observed from Fig. 2, the majority of experimental points either overlap or are located in the vicinity (within the sampling uncertainty) of the central line (I) corresponding to the theoretically expected count median diameters, calculated assuming geometric mean of standard deviation $\sigma_d = \frac{1}{3\sqrt{2}} \ln 10$. Therefore, good agreement between the theory and experimental data can be concluded.

Further analyzing the diagram in Fig. 2, it can be observed that even for the rains with extremely high intensity $I > 100$ mm/h, the expected count median raindrop diameter, $\tilde{d}_n$, remains less than ~6 mm, as given by the line (I). This overlaps with experimental and theoretical observations by Villermaux and Bossa [45], who assumed $We_a = 6$ as the critical Weber number for aerodynamic drop breakup, and calculated the maximum expected raindrop diameter $\langle d \rangle_{max} = \lambda\sqrt{6} = 6.6$ mm, where $\lambda = \sqrt{\gamma/(\rho g)} \approx 2.71$ mm is the capillary length for rainwater. Using Eq. (12) and $\tilde{d}_n = \tilde{d} \exp(-3\sigma_d^2)$, we find that the rain intensity should be as high as $I \approx 56\,000$ mm/h in order for a rain to have $\tilde{d}_n = \langle d \rangle_{max} = 6.6$ mm. The analysis of the data on the world's rainfalls with greatest intensity [86] reveals that the probability of such acute rain is extremely low – there was only one recorded event with rainfall intensity higher than 1000 mm/h ($I = 2280$ mm/h, lasted just for 1 min with overall precipitation 38 mm, recorded in Barot, Guadeloupe on 26 Nov 1970), whereas other 13 recorded extreme rains had lower intensity in the range 76–975 mm/h (see p. 403 [86]). Another observation from line (II) in Fig. 2 is that the smallest possible raindrop diameters in a size distribution, $d_{min}$, are around 0.06 mm at rain intensity 0.05 mm/h and 0.1 mm at 0.4 mm/h, which correspond to the lowest class 1 of drop diameters assessable by laser-optical disdrometers [87] and below the sensitivity of impact disdrometers [88]. The largest drop diameter measurable by laser-optical disdrometers [87] is given by the class 32 with 24.5 mm value, and this value is far beyond the upper boundary of possible maximum raindrop diameters shown by the line (III) in Fig. 2. At the same time, the largest registered raindrops had diameters less than 10 mm, e.g., 9.7 mm observed in Oklahoma [89], 9.2 mm in Japan [90], 8.8 mm in the Marshall Islands [91], 8 mm near Hawaii [92] and others [89]. It can be demonstrated that the possible maximum raindrop diameters in a drop-size distribution, which can be calculated from the balance of work rates of fluid dynamic forces as shown by the line (III) in Fig. 2, can attain values $d_{max} \geq 9.7$ mm for rains with extreme intensities $I \geq 320$ mm/h (this value is beyond the boundaries shown by the line (III) in Fig. 2, so it was calculated using $d_{max} = 4.64\tilde{d}_n$ combined with Eq. (12) and $\tilde{d}_n = \tilde{d} \exp(-3\sigma_d^2)$). Hence the possible maximum raindrop diameters for typical rains $I < 320$ mm/h are normally bounded by the experimentally observed limit of ~10 mm, as illustrated by the line (IV) in Fig. 2. And the absence of experimentally registered raindrop sizes >10 mm even for rains with extreme intensities likely indicates very short lifetimes of such super-large raindrops before disintegration into smaller drops, e.g., $\tau_{burst} < 10$ ms (see Table A1). The breakup can occur spontaneously because of unstable fluid dynamics of drops with $d > 10$ mm, which have Bond (or Eötvös) number greater than 10 and Reynolds number larger than 10,000, so that these raindrops either are short-lived with fluctuating wobbling shapes or burst almost momentarily driven by plausible Rayleigh–Taylor and Kelvin-Helmholtz interface instabilities (see the diagram of shape regimes for bubbles and drops in unhindered gravitational motion through liquids in ref. [93]). Moreover, Komabayasi [40] predicted capillary and gravity waves on the drop surface limit the raindrop diameter to around 8.5 mm. In addition to surface instabilities, for drops with $d > 10$ mm the terminal velocity is very large >15 m/s (see Table A1), so the time scale of collision considerably decreases (see Appendix: Materials and Methods, Eq. (A6)), and respectively the probability of collision and coalescence with smaller raindrops followed by subsequent aerodynamic breakup increases by multiple times. Given that for big diameters the raindrop flight time is on the order of minutes (Table A1), the chance to avoid collisional breakup before reaching the ground for such big raindrops is almost zero. This conclusion is supported by the published experiments on artificial rains [39,40,45], and especially by Blanchard and Spenser [41] who found that initial drops of diameter 10-20 mm falling 60 m at rain intensities 190-1900 mm/h broke up with the final largest diameters ~9 mm observed at the ground.

To extend the comparison with experimental data as shown in Fig. 2, we perform extensive validation between the theoretical count median diameters, respectively $\tilde{d}_n$ (calculated by using Eq. (12) and $\tilde{d}_n = \tilde{d} \exp(-3\sigma_d^2)$) versus the published experimental data, and plot the results in Fig. 3. We examined three different values of the standard deviations $\sigma_d$, corresponding to various widths of the drop-size distribution as discussed above. Figure 3 demonstrates that though in general the value of the standard deviation changes from rain to rain, yet within the sampling uncertainty the experimental points are in the region between two lines with $\sigma_d = \frac{1}{6}\ln 10$ and $\sigma_d = \frac{1}{3}\ln 10$. Therefore, our previous conclusions drawn from the theoretical analysis and Fig. 2, which suggested that $\frac{1}{6}\ln 10 \leq \sigma_d \leq \frac{1}{3}\ln 10$ and $4\sigma_d \leq \ln\frac{d_{max}}{d_{min}} \leq 8\sigma_d$ for a steady rain with insignificant evaporation, are supported by the experimental data obtained by different research groups at various ground locations around the world. Moreover, in contrast to prior research that frequently fitted the observed raindrops size distributions by the lognormal function with the standard deviation $\sigma_d$ as an empirical parameter [75,76,80], here we established that $\sigma_d$ is a measure of the energy content of the stochastic fluctuations of the energy of drops in a rain (see Eqs. (11) and (13)). Further research is required to determine how the standard deviation can be calculated more accurately as a function of rain intensity, type of rain, pattern of air flow, and other atmospheric conditions.

Next, we present a comparison between the calculated and measured mass-weighted and number-weighted time-averaged raindrop size distributions for rains with different intensities, as shown in Fig. 4. The values of the coefficient of determination, $R^2$, indicating the proportion of the variation in the dependent variable that is predictable from the independent variable in a regression analysis [94], serve as a quantitative measure of the agreement between the calculated and experimental points. Figure 4 demonstrates that on average, $R^2 > 0.95$. Given the measurement uncertainties (shown as error bars in Fig. 4) and drop-size range truncation and bias found in the measuring instruments (not reflected in Fig. 4 yet extensively reported elsewhere [82,83,95,75]), the observed values of $R^2$ indicate an overlap between the theoretical and the experimental size distributions.

The Supplementary Information to this paper contains the table summarizing the measured count median diameters of raindrops extracted from the published experimental rain spectrograms (Table S1), and additional figures, demonstrating the overlap between theoretical and experimental drop size distributions for rains with various intensity (Figs. S1-S13), and theoretical diagram of mass median raindrop diameters as a function of rain intensity (Fig. S14).

### B. Concentration of raindrops at ground level

To enhance the comprehensive comparison between our theory and published experimental observations of drop-size distributions and expected raindrop diameters across a wide range of rain intensities (0.4-40 mm/h), we have also undertaken the task of calculating the average concentration of raindrops in the air at ground level. According to Marshall and Palmer [73], who analyzed their own and Law and Parson's [74] experimental data, the number concentration of raindrops in air can be fitted by the following exponential law:

$$N(d) = N_0 e^{-\Lambda d}, \tag{15}$$

where $N(d)\Delta d$ is number of raindrops with diameter between $d$ and $d+\Delta d$ in unit volume of space, $N_0 = 0.08$ cm$^{-4}$ is value of $N(d)$ for $d \to 0$, and $\Lambda = 41 I^{-0.21}$ with $I$ in mm/h and $\Lambda$ in cm$^{-1}$. The constant $N_0$ can be interpreted as the average number concentration (spatial density) of raindrops at the ground level [45] per unit of drop size. On the other hand, the rain intensity is connected with the number concentration of raindrops by the expression [45,96,97]:

$$I = \int_0^\infty N(d) \frac{\pi d^3}{6} U(d) \, \mathrm{d}d, \tag{16}$$

where $U(d)$ is raindrop free-fall velocity. Combining Eqs. (15) and (16), and considering $U(d) = C\sqrt{(\rho/\rho_a)gd}$, where $C = 1.0 - 1.6$ is a constant accounting for the drag coefficient and shape of a falling raindrop [36,45,98,99], we find:

$$I = \frac{35\pi\sqrt{\pi}C}{32}\sqrt{\frac{\rho}{\rho_a}g}\frac{N_0}{\Lambda^{9/2}}, \tag{17}$$

Next, we consider the mode (most frequent) diameter of the count-weighted raindrop size distribution as $\hat{d}_n = 1/\Lambda$, and combine Eqs. (12) and (17) with $\hat{d}_n = \tilde{d}_n \exp(-\sigma_d^2)$ and $\tilde{d} = \tilde{d}_n \exp(3\sigma_d^2)$ [64,65] to obtain:

$$N_0 = \frac{32}{105\sqrt{\pi}C\hat{A}\lambda^2}\sqrt{\frac{\rho_a}{\rho}}\exp\left(\frac{39}{2}\sigma_d^2 + 1\right) \tag{18}$$

From here, by using the average value $C = 1.3$ and by choosing the value $\sigma_d = 0.66$ out of the established range $\frac{1}{6}\ln 10 \leq \sigma_d \leq \frac{1}{3}\ln 10$, we find $N_0 \approx 0.08$ cm$^{-4}$, which is consistent with Marshall and Palmer's experimental findings [73].

The total number concentration of raindrops in air, $N_T$, according to Feingold and Levin [75] is given by the expression $N_T = 172 I^{0.22}$, and the number-weighted raindrop diameter is determined by $\tilde{d}_n = 0.75 I^{0.21}$ (here $N_T$ has units of 1/m³, $I$ is in mm/h and $\tilde{d}_n$ is in mm). Correspondingly, $N_T = 323 \cdot 10^3 \tilde{d}_n^{1.05}$, where the diameter is in meters. Feingold and Levin [75] also found that the number concentration of raindrops of diameter $d$ can be described by the lognormal function:

$$N(d) = \frac{N_T}{d\,\sigma_d\sqrt{2\pi}} \exp\left[-\frac{1}{2}\left(\frac{\ln d/\tilde{d}_n}{\sigma_d}\right)^2\right] \tag{19}$$

Substituting Eq. (19) into (16), then using $U(d) = C\sqrt{(\rho/\rho_a)gd}$ for the integration, and after that combining the result with Eqs. (12) and $\tilde{d} = \tilde{d}_n \exp(3\sigma_d^2)$, leads to:

$$N_T = \frac{2\tilde{d}_n}{C\hat{A}\lambda^2}\sqrt{\frac{\rho_a}{\rho}} \exp\left(\frac{71}{8}\sigma_d^2 + 1\right) \tag{20}$$

The above equation provides the overlap with Feingold and Levin's [75] finding $N_T = 323 \cdot 10^3 \tilde{d}_n^{1.05} \approx 255 \cdot 10^3 \tilde{d}_n$ for $C = 1.3$ and $\sigma_d = 0.537 \approx 0.54$. The latter value is consistent with the standard deviation $\sigma_d = \frac{1}{3\sqrt{2}}\ln 10 \approx 0.54$ that we used to construct the diagram of expected raindrop diameters in Fig. 2.

Analyzing Eqs. (18) and (20), we conclude that both $N_0$ and $N_T$ depend on the surface tension and density of rainwater, density of air, gravitational acceleration and magnitude of stochastic energy fluctuations. Therefore, $N_0$ and $N_T$ change with the atmospheric and planetary conditions, in particular atmospheric temperature, pressure, humidity, turbulence, radiation level, wind and vary with altitude above the ground. These findings are in qualitative agreement with the observations [76,100–103].

## IV.    SUMMARY AND CONCLUSION

In this work we developed a theory based on thermodynamics and fluid dynamics to determine the expected median diameter and stationary size distribution of raindrops at the ground as a function of rain intensity. First, we examined the characteristic time scales and performed dimensional and scale analyses of physical phenomena and processes occurring during the fall of raindrops. We found that compared to typical collision time scale, the time scales of gravity, aerodynamic drag, drop bursting and capillary phenomena are smaller by two or even three orders of magnitude (see Table A1). We deduced that these micro-processes normally provide the main contribution to the droplet size distribution observed at the ground. Furthermore, we introduced the rain frequency as an intensive parameter describing the rate at which raindrops land on the ground. This enabled us to derive the characteristic non-dimensional equation of the balance of specific work rates acting on a falling raindrop (Eq. (4)), establish governing dimensionless numbers and derive the underlying equation of the expected raindrop diameters (Eq. (5)). Applying the principles of possible, impossible and expected drop diameters, which we described recently for aerosols and sprays [61], we constructed the diagram of expected raindrop diameters (Fig. 2).

Next, we developed the theory of the size distribution of raindrops using the deterministic mass, momentum, and energy balances combined with a stochastic approach to address the energy fluctuations in rainfall, as generally given by the system of equations (1), (2), (10), (A11) and (A12). We considered the work rates of external forces acting on a falling raindrop, including aerodynamic drag and the force of Earth's gravity field, to be employed against the surface tension and viscosity of raindrop's water (Eqs. (1) and (2)). Together with random events accounted for by the stochastic fluctuation term in Eq. (10), we deduced these interactions are responsible for the evolution of the size of raindrops travelling from the clouds to the ground. The stochastic fluctuations are the random disturbances emerging from both the raindrops and the surrounding atmospheric air flow, and can be caused by bouncing, breakup and coalescence of colliding raindrops, drop oscillations, internal circulation, vortex shedding, air flow turbulence and other stochastic atmospheric effects. This theoretical framework led to the derivation of the central stochastic differential equation (11), describing the distribution of raindrop masses at various conditions. Analyzing the central stochastic differential equation (11), we established the analytical dependence of mass median raindrop diameter on rain intensity (Eq. (12)). The general solution of the central stochastic differential equation (11) provides a family of functions describing the distribution of raindrop masses at various conditions. We showed that the lognormal drop-size distribution (given by Eqs. (13) and (14)) for mass- and number-weighted probability density functions, respectively) is a particular solution obtained for a steady rain and white noise energy fluctuations.

The results of extensive comparison between the theory and the published experimental data demonstrated a good agreement between the theoretically predicted and observed median diameters and size distribution functions of raindrops (Figs. 2-4, and Figs. S1-S14 in the Supplementary Information). In addition, consistent with the previously reported experimental observations on super-large raindrops, our theory predicted that possible maximum raindrop diameters for typical rains with intensity <320 mm/h are normally bounded by the limit of ~10 mm imposed by the interplay between the aerodynamic drag and gravity forces acting against the surface tension and viscous forces (Fig. 2). And for rare rainfalls with intensity >320 mm/h, the elevated drop surface instabilities and amplified frequency of inter-raindrop collisions are responsible for the rapid breakup of raindrops with diameters >10 mm, so such raindrops, if ever formed, are extremely short-lived and are not observed at the ground.

From the developed theory we also established analytical expressions for the Marshall-Palmer's constant (Eq. (18)) and the total number concentration of raindrops in air (Eq. (20)). These formulas exhibit overlap with experimental values, demonstrating that both parameters are contingent upon factors such as surface tension and density of rainwater, air density, gravitational acceleration, and the magnitude of stochastic energy fluctuations. Moreover, the total number concentration reveals a correlation with the median raindrop diameter, making it dependent on rainfall intensity.

Future research can center on unraveling the influence of evaporation on raindrop size distribution, with a specific emphasis on its proclivity to shift the drop-size distribution towards smaller raindrop sizes. Also, there is a need to refine the calculation of the standard deviation of raindrop size distribution. This step should encompass comprehensive consideration of factors such as rain intensity, rain type, airflow patterns, and atmospheric conditions to enhance accuracy in assessing raindrop size variability.

In conclusion, the developed theory and its broader theoretical approach have potentially far-reaching applications. Beyond rain, they can be utilized in climate models and related fields, including liquid atomization of sprays and aerosols, bubble and particle production, planetary science for forecasting drop-size distributions of non-water precipitation on other celestial bodies, and in geophysics to assess Earth's conditions in ancient eras.

**ACKNOWLEDGMENTS**

The research reported in this publication was supported by the Foundation for Health Advancement and New Jersey Health Foundation via the Allergan Foundation Innovation Grant Program, Grant # ALL 01-21, and the National Center for Advancing Translational Sciences (NCATS), a component of the National Institute of Health (NIH) under award number UL1TR003017. The content is solely the responsibility of the authors and does not represent the official views of the National Institutes of Health. The authors also acknowledge the support from the University City Science Center via QED Proof-of-Concept Program (grant number QED 2102) and the support from Princeton University via Intellectual Property Accelerator Fund program.

**APPENDIX: MATERIALS AND METHODS**

1. **Theoretical developments**

   a. *Time scales*

By scrutinizing the individual time scales associated with various physical phenomena and processes occurring during the fall of raindrops, we can enhance our understanding of their respective rates. The time scales of physical phenomena are capillary (surface tension), $\tau_\gamma$, viscosity, $\tau_\nu$, gravity, $\tau_g$, and drag, $\tau_D$:

$$\tau_\gamma \sim \left(\frac{\rho \ell^3}{\gamma}\right)^{1/2}, \quad \tau_\nu \sim \frac{\ell^2}{\nu}, \quad \tau_g \sim \left(\frac{\ell}{g}\right)^{1/2}, \quad \tau_D \sim \left(\frac{m\ell}{F_D}\right)^{1/2}, \tag{A1}$$

where $\ell$ is characteristic length scale, $g$ is gravity acceleration, $m$ is drop mass, $F_D$ is drag force, and $\rho$, $\gamma$, $\nu$ are rainwater density, surface tension and kinematic viscosity, respectively. Assuming raindrop motion at terminal velocity so that drag and gravity forces are equal, $F_D = mg$, we conclude $\tau_g = \tau_D$. The process time scales are particle relaxation time scale [64], $\tau_p$, time to reach the terminal velocity [64], $\tau_t$, and raindrop flight time, $\tau_f$:

$$\tau_p = \frac{V_t}{g}, \quad \tau_t = 3\tau_p, \quad \tau_f = \tau_t + \frac{z_i - \Delta z_t}{V_t}, \tag{A2}$$

where $V_t$ is raindrop terminal velocity, $z_i$ is the initial height of a raindrop falling after formation in clouds, and $\Delta z_t$ is vertical distance travelled by a raindrop before it reaches the terminal velocity [64]:

$$\Delta z_t = z_i - z(\tau_t) = V_t \tau_t - (V_t - V_0)\tau_p \left(1 - e^{-\tau_t/\tau_p}\right), \tag{A3}$$

By assuming the initial velocity of a raindrop after it departs from the clouds is close to zero, $V_0 \approx 0$, and noting that $1 - e^{-\tau_t/\tau_p} = 1 - e^{-3} = 0.95$, we find:

$$\Delta z_t = 3 V_t \tau_p - 0.95 V_t \tau_p = 2.05 V_t \tau_p, \tag{A4}$$

The time scale of aerodynamic bursting of a raindrop having diameter, $d$, is calculated using an equation proposed by Villermaux and Bossa [45]:

$$\tau_{burst} = \frac{d}{2V_t}\sqrt{\frac{\rho}{\rho_a(1 - 6/\mathrm{We}_a)}}, \tag{A5}$$

where $\rho_a$ is air density and $\mathrm{We}_a = \rho_a V_t^2 d / \gamma$ is the Weber number based on air density. Note that Eq. (A5) assumes $\mathrm{We}_a > 6$ as a necessary condition for aerodynamic breakup of a raindrop, which is similar to the observed in the secondary atomization of spray droplets [104]. In addition, the process of raindrop fall involves inter-drop collisions, and the respective time scale is approximated by [45]:

$$\tau_{coll} = \frac{d}{N_0 \tilde{d}^4 V_t}, \tag{A6}$$

where $\tilde{d}$ is the mean diameter of raindrops obtained from initial drops with diameter $d$ as the result of collisions, and $N_0 = 0.08 \text{ cm}^{-4}$ is Marshall-Palmer's constant corresponding to the average spatial density of raindrops observed for rains of different intensity at the ground level [73]. For typical mean raindrop diameters $\tilde{d} = 1-3 \text{ mm}$, by assuming initial parental drops with $We_a \gg 6$ (see ref. [45]), we find from Eqs. (A5) and (A6) that $\tau_{coll} = \tau_{burst} \frac{2}{N_0 \tilde{d}^4} \sqrt{\frac{\rho_a}{\rho}} \approx (100-10,000) \tau_{burst}$. By setting $\tau_{burst} = 0.01 \text{ s}$ (see Table A1), we obtain the range of inter-drop collision time scales $\tau_{coll} = 1-100 \text{ s}$, which is consistent with the previous literature reports [51,105–108].

As a first approximation, we assume that characteristic length scale in Eq. (A1) is equal to the drop diameter, $\ell = d$. Using rainwater properties $\rho = 1000 \text{ kg/m}^3$, $\gamma = 0.072 \text{ N/m}$, $\mu = 0.001 \text{ Pa} \cdot \text{s}$, $\nu = \mu/\rho = 10^{-6} \text{ m}^2/\text{s}$ and the air density $\rho_a = 1.2 \text{ kg/m}^3$, we apply Eq. (A1) to calculate the values of the phenomena time scales for different raindrop diameters. The height of clouds with raindrops are typically $O(1\text{-}10 \text{ km})$ [37], so we take a conservative value $z_i = 1000 \text{ m}$ for the time of flight calculation. The acceleration of gravity is assumed to be constant for the whole raindrop flight, $g = 9.81 \text{ m}^2/\text{s}$. The summary of the calculated time scales for four representative raindrop diameters is given in Table A1.

TABLE A1. Calculated time scales of physical phenomena and processes involved in raindrop fall. The capillary (surface tension), $\tau_\gamma$, viscosity, $\tau_\nu$, gravity, $\tau_g$, drag, $\tau_D$, particle relaxation, $\tau_p$, time to reach the terminal velocity, $\tau_t$, and raindrop flight time, $\tau_f$, characteristic times are given for drops of different diameters falling from initial height $z_i = 1000 \text{ m}$ in quiescent air. For the reference, collision time scales calculated by Eq. (A6) $\tau_{coll} = 1-100 \text{ s}$.

| d, mm | $\tau_\gamma \cdot 10^3$, s | $\tau_\nu$, s | $\tau_g = \tau_D$ $\cdot 10^3$, s | $V_t$, m/s | $We_a$ | $\tau_p$, s | $\tau_t$, s | $\Delta z_t$, m | $\tau_{burst} \cdot 10^3$, s | $\tau_f$, s |
|---|---|---|---|---|---|---|---|---|---|---|
| 0.1 | 0.1 | 0.01 | 3 | 0.29 | $1.4 \cdot 10^{-4}$ | 0.03 | 0.09 | 0.02 | - | 3395 |
| 1 | 4 | 1 | 10 | 3.92 | 0.26 | 0.40 | 1.20 | 3.21 | - | 255 |
| 3 | 19 | 9 | 18 | 8.08 | 3.27 | 0.82 | 2.47 | 13.7 | - | 125 |
| 5 | 42 | 25 | 23 | 9.08 | 6.87 | 0.93 | 2.78 | 17.2 | 22 | 111 |
| 10 | 118 | 100 | 32 | 15.6 | 40.4 | 1.59 | 4.76 | 50.7 | 10 | 66 |

*The terminal velocities, $V_t$, are calculated based on the experimental data of Gunn and Kinzer [109].

The analysis of Table A1 leads to conclusion that the time scale associated with viscosity of raindrops is much larger than the time scales of capillary, gravity, and drag phenomena. Hence, the role of rainwater viscosity in the process of establishing the size distribution of raindrops is negligible. Furthermore, whereas the time scales of gravity and drag are substantially larger than the time scale of surface tension for diameters smaller than 1 mm, for a raindrop with ~3 mm diameter these timescales become equal, and for bigger raindrops >3 mm in diameter the gravity and drag effects occur on shorter time scales than the surface tension phenomena. Also, the gravity and drag influence raindrops on a considerably shorter time scale than the particle relaxation time and time to reach the terminal velocity, which may suggest that breakup of big raindrops is possible even before they reach terminal velocity. Compared to the calculated above typical collision time scales 1-100 s, the time scales of gravity, drag, bursting and capillary phenomena are smaller by two or even three orders of magnitude. This implies that these micro-processes provide the main contribution to the droplet size distribution observed at the ground. At the same time, the process of inter-drop collisions cannot be neglected completely, because the typical time of raindrop flight is on the order of hundreds of seconds, which is greater than the typical collision time between raindrops. In addition, our estimation of the collision time scale by Eq. (A6) is based on the value of average spatial density of raindrops observed at the ground, $N_0 = 0.08 \text{ cm}^{-4}$. The actual spatial drop density can vary with rain intensity and can be higher near the cloud base than at the ground. Therefore, under some conditions probably ones with high rain intensity and at high altitudes closer to the cloud base, the collision time scale can compete with gravity, drag, capillary and bursting time scales, and thus play a dominant role in the observed raindrop size distributions. In the current study, we do not consider such extreme conditions, and account for inter-drop collisions as contributing to the stochastic fluctuation term in the energy balance, as given by Eq. (10).

b. *Scale analysis*

The energy scale of a physical phenomenon, $e$, can be considered as the ratio $e = \dfrac{\ell^2}{\tau^2}$, where $\ell$ and $\tau$ are length and time scales associated with the phenomenon, respectively. Correspondingly, using Eq. (A1), the energy scales of capillary (surface tension), $e_c$, viscosity, $e_v$, gravity, $e_g$, and drag phenomena, $e_D$, are given by:

$$e_\gamma = \frac{\ell^2}{\tau_\gamma^2} \sim \frac{\gamma}{\rho\ell}, \quad e_v = \frac{\ell^2}{\tau_v^2} \sim \frac{v^2}{\ell^2}, \quad e_g = \frac{\ell^2}{\tau_g^2} \sim g\ell, \quad e_D = \frac{\ell^2}{\tau_D^2} \sim \frac{F_D \ell}{m}. \tag{A7}$$

Assuming that at the first approximation $\ell = d$, we can determine the following set of dimensionless numbers, which reflect six different ratios of the energy scales:

$$\frac{e_v}{e_\gamma} \sim \frac{\mu^2}{\rho\gamma d}, \quad \frac{e_g}{e_\gamma} \sim \frac{\rho g d^2}{\gamma}, \quad \frac{e_D}{e_\gamma} \sim \frac{F_D \rho \ell^2}{m\gamma}, \quad \frac{e_g}{e_v} \sim \frac{\rho^2 g d^3}{\mu^2}, \quad \frac{e_D}{e_v} \sim \frac{F_D \rho^2 \ell^3}{m\mu^2},$$
$$\frac{e_D}{e_g} \sim \frac{F_D}{mg} \tag{A8}$$

Here we can recognize Ohnesorge number, $\text{Oh} = \dfrac{\mu}{\sqrt{\rho\gamma d}}$, Bond / Eötvös number $\text{Bo} = \dfrac{\rho g d^2}{\gamma}$ and Galilei number (or Archimedes number) $\text{Ga} = \dfrac{\rho^2 g d^3}{\mu^2}$. On the other hand, the ratios of energy scales can be considered proportional to the ratios of the respective specific works, $\dfrac{e_v}{e_\gamma} \sim \dfrac{w_v}{w_\gamma}, \dfrac{e_g}{e_\gamma} \sim \dfrac{w_g}{w_\gamma}$ and so on for other ratios. Thus, from Eq. (A8) we can derive $\dfrac{w_v}{w_\gamma} \propto \dfrac{\mu^2}{\rho\gamma d}, \dfrac{w_g}{w_\gamma} \propto \dfrac{\rho g d^2}{\gamma}, \dfrac{w_g}{w_v} \propto \dfrac{\rho^2 g d^3}{\mu^2}$ and so on for other ratios of specific work rates. Therefore, $\dfrac{w_v}{w_\gamma} \propto \text{Oh}^2, \dfrac{w_g}{w_\gamma} \propto \text{Bo}$ and $\dfrac{w_g}{w_v} \propto \text{Ga}$.

The scale of energy rate of a phenomenon, $\varepsilon$, can be determined as the ratio $\varepsilon = \dfrac{e}{\tau} = \dfrac{\ell^2}{\tau^3}$, Correspondingly, the scales of energy rates of capillary (surface tension), $\varepsilon_c$, viscosity, $\varepsilon_v$, gravity, $\varepsilon_g$, and drag phenomena, $\varepsilon_D$, are given by:

$$\varepsilon_\gamma = \frac{e_\gamma}{\tau_R} \sim \frac{\gamma \omega_R}{\rho \ell}, \quad \varepsilon_v = \frac{e_v}{\tau_v} \sim \frac{v^3}{\ell^4}, \quad \varepsilon_g = \frac{e_g}{\tau_g} \sim g^{3/2}\ell^{1/2}, \quad \varepsilon_D = \frac{e_D}{\tau_D} \sim \frac{F_D^{3/2}\ell^{1/2}}{m^{3/2}}. \tag{A9}$$

As a first approximation, we assume as above $\ell = d$. From Eq. (A9) we derive the following set of dimensionless numbers, which reflect six ratios of the energy rate scales:

$$\frac{\varepsilon_v}{\varepsilon_\gamma} \sim \frac{\mu^3}{\gamma \rho^2 d^3 \omega_R}, \quad \frac{\varepsilon_g}{\varepsilon_\gamma} \sim \frac{\rho g^{3/2} d^{3/2}}{\gamma \omega_R}, \quad \frac{\varepsilon_D}{\varepsilon_\gamma} \sim \frac{F_D^{3/2} \rho d^{3/2}}{m^{3/2}\gamma \omega_R}, \quad \frac{\varepsilon_g}{\varepsilon_v} \sim \frac{\rho^3 g^{3/2} d^{9/2}}{\mu^3},$$

$$\frac{\varepsilon_D}{\varepsilon_v} \sim \frac{F_D^{3/2}\rho^3 d^{9/2}}{m^{3/2}\mu^3}, \quad \frac{\varepsilon_D}{\varepsilon_g} \sim \frac{F_D^{3/2}}{m^{3/2}g^{3/2}}. \tag{A10}$$

On the other hand, the ratios of energy rate scales can be considered proportional to the ratios of the respective rates of specific works, $\dfrac{\varepsilon_v}{\varepsilon_\gamma} \sim \dfrac{\dot{w}_v}{\dot{w}_\gamma}$, $\dfrac{\varepsilon_g}{\varepsilon_\gamma} \sim \dfrac{\dot{w}_g}{\dot{w}_\gamma}$ and so on for other ratios of energy rate scales. Thus, from Eq. (A10) we can derive $\dfrac{\dot{w}_v}{\dot{w}_\gamma} \propto \dfrac{\mu^3}{\gamma \rho^2 d^3 \omega_R}$, $\dfrac{\dot{w}_g}{\dot{w}_\gamma} \propto \dfrac{\rho g^{3/2} d^{3/2}}{\gamma \omega_R}$ and so on for other ratios of specific work rates.

### c. Size distribution of raindrops

The process of raindrops flowing from the clouds to the ground is analyzed by applying the law of mass conservation and the first law of thermodynamics for a control volume including both liquid and gas phases, as shown in Fig. 1c. The rate of change of mass, $m_{cv}$, in the control volume is determined by the difference of mass flow rates at the inlet, $\dot{m}_i$, and exit, $\dot{m}_e$:

$$\frac{dm_{cv}}{dt} = \dot{m}_i - \dot{m}_e \tag{A11}$$

The rate of change of the number of drops in the control volume, $n_{cv}$, is given by the difference of the net rate of drops at the inlet, $\dot{n}_i$, and exit, $\dot{n}_e$, plus source term, $\dot{s}$, expressing the rate of change due to drop breakup or coalescence events inside the control volume:

$$\frac{dn_{cv}}{dt} = \dot{n}_i - \dot{n}_e + \dot{s} \tag{A12}$$

Performing the energy balance for the flow of raindrops (see Fig. 1c) and accounting for stochastic energy fluctuations, leads to the underlying differential equation of energy balance as given by Eq. (10). Next, the system of equations (1), (2), (10), (A11) and (A12) can be rendered to the forms typical in multiphase transport phenomena by dividing those equations with the total volume of the control volume to express the balances for mass, number and energy fractions for the dispersed phase in two-phase flow of raindrops and air. However, this step is not necessary, because in our analytical solution we will further transform those equations to express the energy balance in terms of ratios of mass, energy and work. Expanding previously obtained reduced form of Eq. (10), $\dot{E} = \dot{W} + A_\xi \xi(t)$, we get:

$$\frac{d}{dt}(nme_d) = -\frac{d}{dt}\left[n(W_\gamma + W_\nu)\right] + A_\xi \xi(t) \tag{A13}$$

Here we used $E = nme_d$, where $n$ is number of raindrops in control volume, $e_d$ is specific energy of raindrop including internal energy and kinetic and potential energy components. In addition, in order to develop Eq. (A13), similarly to Eq. (2) we considered the work done on raindrops, $W$, to be equal to the minus sum of the work done by the drop surface tension, $W_\gamma$, and viscous energy dissipation, $W_\nu$, i.e., $W = -(W_\gamma + W_\nu)$, where the minus sign is to account for the sign convention for work in thermodynamics [62]. Combining the equations (A11)-(A13), and expanding the terms to explicitly identify the dependence on the raindrop mass, we obtain:

$$\frac{1}{m}\frac{dm}{dt} = -\frac{\dot{n}}{n}\left[1 + \frac{1}{\alpha}\left(1 + \frac{w_\nu}{w_\gamma}\right) + \frac{1}{\beta}\left(1 + \frac{\dot{w}_\nu}{\dot{w}_\gamma}\right)\right] + \frac{A_\xi}{mne_d}\xi(t), \tag{A14}$$

where $\dot{n} = \frac{dn}{dt}$, $\dot{w}_\gamma = \frac{\dot{W}_\gamma}{m}$, $\dot{w}_\nu = \frac{\dot{W}_\nu}{m}$, $w_\gamma = \frac{W_\gamma}{m}$, $w_\nu = \frac{W_\nu}{m}$, and $\alpha = \frac{e_d}{w_\gamma}$, $\beta = \frac{e_d}{\dot{w}_\gamma}\frac{\dot{n}}{n}$. Next, we employ the results of dimensional and scale analyses, $\frac{\dot{w}_g}{\dot{w}_\gamma} \propto \frac{\rho g^{3/2} d^{3/2}}{\gamma \omega_R}$, and Eq. (3) to establish $1 + \frac{\dot{w}_\nu}{\dot{w}_\gamma} = C_1 \cdot 2\frac{\rho g^{3/2} d^{3/2}}{\gamma \omega_R}$, where $C_1$ is a scaling factor for the balance of ratios of rates of specific works. In similar way, we obtain $1 + \frac{w_\nu}{w_\gamma} = C_2 \cdot 2\text{Bo}$, where Bo is Bond number that we determined in the previous section, and $C_2$ is a scaling factor for the balance of ratios of specific works. On the other hand, we established above that $\frac{w_\nu}{w_\gamma} \propto \text{Oh}^2$, and for drops of rainwater for the possible sizes in rain $d = 0.1-10$ mm (see Fig. 2), we find that $\text{Oh}^2 = 10^{-6} - 10^{-4}$. Correspondingly, using Taylor expansion $e^{w_\nu/w_c} = 1 + \frac{w_\nu}{w_c} + O\left(\left(\frac{w_\nu}{w_c}\right)^2\right)$, and noting that $O\left(\left(\frac{w_\nu}{w_c}\right)^2\right) = O(\text{Oh}^4) \ll O\left(\frac{w_\nu}{w_c}\right)$, we conclude that $e^{w_\nu/w_c} \simeq 1 + \frac{w_\nu}{w_c}$. Therefore, $\frac{w_\nu}{w_c} \simeq \ln\left(1 + \frac{w_\nu}{w_c}\right)$ and thus $1 + \frac{w_\nu}{w_c} \simeq 1 + \ln(C_2 \cdot 2\text{Bo})$. In similar way, we obtain that $1 + \frac{\dot{w}_\nu}{\dot{w}_c} \simeq 1 + \ln\left(C_1 \cdot 2\frac{\rho g^{3/2} d^{3/2}}{\gamma \omega_R}\right)$. Substitution the last two expressions into Eq. (A14) and performing basic algebraic manipulations, we get:

$$\frac{1}{m}\frac{dm}{dt} = -\frac{\dot{n}}{n}\ln\left[e^{(1+\ln C_2)/\alpha + (1+\ln C_1)/\beta + 1}(2\text{Bo})^{1/\alpha}\left(2\frac{\rho g^{3/2} d^{3/2}}{\gamma \omega_R}\right)^{1/\beta}\right] + \frac{A_\xi}{mne_d}\xi(t), \tag{A15}$$

By utilizing $d = \left(\dfrac{6m}{\pi\rho}\right)^{1/3}$ and further analyzing and transforming the right hand side of Eq. (A15), we derive the stochastic differential equation (11). Afterwards, our analysis leads us to the conclusion that Eq. (11) characterizes a stochastic process described by Uhlenbeck and Ornstein [77] with a random variable $x = \ln\dfrac{m}{\tilde{m}}$, where $\tilde{m}$ is the median mass of raindrops. Introducing a mass-weighted probability density function, $q_m(x,t)$, such that $\int_0^\infty q_m(x,t)dx = 1$, and assuming that stochastic fluctuations in rain $\sigma\xi(t)$ can be approximated by a white noise, Eq. (11) can be transformed into a Fokker-Planck partial differential equation [78]:

$$\frac{\partial q_m}{\partial t} = a\frac{\partial}{\partial x}(xq_m) + \frac{\sigma^2}{2}\frac{\partial^2 q_m}{\partial x^2}, \tag{A16}$$

Considering steady rainfall, $\dfrac{\partial q_m}{\partial t} = 0$. The stationary solution of Eq. (A16) leads to a normal distribution of the probability density function $q_m\left(x = \ln\dfrac{m}{\tilde{m}}\right)$ describing the likelihood of a raindrop to have mass between $x = \ln\dfrac{m}{\tilde{m}}$ and $x + \Delta x = \ln\dfrac{m}{\tilde{m}} + \Delta\ln\dfrac{m}{\tilde{m}}$:

$$q_m\left(\ln\frac{m}{\tilde{m}}\right) = \frac{1}{\sigma_1\sqrt{2\pi}}\exp\left[-\frac{1}{2}\left(\frac{\ln(m/\tilde{m})}{\sigma_1}\right)^2\right], \tag{A17}$$

where $\sigma_1 = \sigma/\sqrt{2a}$. Using the relation between mass and diameter of a spherical drop, $\ln\dfrac{m}{\tilde{m}} = 3\ln\dfrac{d}{\tilde{d}}$, where $\tilde{d} = \left(\dfrac{6\tilde{m}}{\pi\rho}\right)^{1/3}$ is the mass median diameter of raindrops, the stochastic differential equation (11) can be rewritten in terms of raindrop diameter:

$$\frac{d}{dt}\left(\ln\frac{d}{\tilde{d}}\right) = -a\left(\ln\frac{d}{\tilde{d}}\right) + \frac{\sigma}{3}\xi(t) \tag{A18}$$

The corresponding Fokker-Plank equation is given by Eq. (A16) for the probability density function $q_m\left(\ln\dfrac{d}{\tilde{d}}, t\right)$. In a steady state, $\dfrac{\partial q_m}{\partial t} = 0$, and thus the stationary solution is given by a normal distribution function of logarithm of raindrop diameter, $q_m\left(\ln\dfrac{d}{\tilde{d}}\right)$:

$$q_m\left(\ln\frac{d}{\tilde{d}}\right) = \frac{1}{\sigma_d\sqrt{2\pi}}\exp\left[-\frac{1}{2}\left(\frac{\ln(d/\tilde{d})}{\sigma_d}\right)^2\right], \tag{A19}$$

where $\sigma_d = \sigma/(3\sqrt{2a})$ is a standard deviation of the logarithm of drop diameters. The mass median diameter of raindrops is such that for $d = \tilde{d}$, there is the relationship $\int_{-\infty}^{0} q_m(x)dx = 0.5$, where $x = \ln\frac{d}{\tilde{d}}$. The Eq. (A19) results in a lognormal form of the mass-weighted probability density function of raindrop diameter, $q_m(d)$, as given by Eq. (13). Introducing a number-weighted probability density function, $q_n\left(x = \ln\frac{d}{\tilde{d}_n}\right)$, where $\int_{-\infty}^{+\infty} q_n(x)dx = 1$ and $\tilde{d}_n$ is count median diameter such that for $d = \tilde{d}_n$, there is $\int_{-\infty}^{0} q_n(x)dx = 0.5$, it can be shown that [64]:

$$q_n\left(\ln\frac{d}{\tilde{d}_n}\right) = \frac{1}{\sigma_d\sqrt{2\pi}}\exp\left[-\frac{1}{2}\left(\frac{\ln(d/\tilde{d}_n)}{\sigma_d}\right)^2\right] \qquad (A20)$$

The connection between mass and count median diameters is given by Hatch-Choate expression $\tilde{d}_n = \tilde{d}\exp(-3\sigma_d^2)$ [64,65]. The Eq. (A20) results in a lognormal form of the mass-weighted probability density function of raindrop diameter, $q_n(d)$, as given by Eq. (14).

To determine the standard deviation of the probability density function, $\sigma_d$, we employ so-called "sigma rules" of a normal distribution function [81]. The idea is that according to Eq. (A20), the probability density function of the logarithm of drop diameter, $q_n(\ln d/\tilde{d}_n)$, follows a normal Gauss-Laplace distribution with the same standard deviation $\sigma_d$. Therefore, 95.45% of the raindrops have the diameters in the range of $-2\sigma_d \leq \ln(d/\tilde{d}_n) \leq 2\sigma_d$, and 99.99% of raindrops have the diameters in the range of $-4\sigma_d \leq \ln(d/\tilde{d}_n) \leq 4\sigma_d$ [110]. Then, it is reasonable to expect that the logarithmic width of the raindrop size distribution observed experimentally, $\ln\frac{d_{max}}{d_{min}}$, is between $4\sigma_d$ and $8\sigma_d$. On the other hand, we have previously found that $d_{max}/d_{max} = 10^{4/3}$. Therefore, we expect that the values of the standard deviation are confined by the range $\frac{1}{6}\ln 10 \leq \sigma_d \leq \frac{1}{3}\ln 10$, and we test this conclusion versus the published experimental data as given in Fig. 3. Using the geometric mean value of $\sigma_d = \frac{1}{3\sqrt{2}}\ln 10 \approx 0.54$, we plotted the diagram of raindrop diameters in Fig. 2 with respect to rain intensities. The latter diagram can be used to obtain the possible range and the expectation of raindrop diameters for a given intensity of rainfall.

Based on the good agreement between the theory and observations, we deduce that in many rainfalls the time-averaged drop-size distributions observed on the ground can be well-described by considering steady state and stochastic fluctuations in rain approximated by white noise. Furthermore, it is worth noting that though the majority of the published experimental reports lack information on the air humidity during the rainfall, the evaporation of raindrops shifts the droplet size distribution towards the smaller drop diameters and thus for such rains the characteristic raindrop diameters (e.g., mass and count median, mass and count mean) are smaller than those for rains of the same intensity with negligible evaporation. On the other hand, the addition of evaporation terms into the system of mass and energy conservation equations has no influence on the basic form of the underlying stochastic differential equation (Eq. (11)), yet it changes the coefficients $a$ and $\sigma$, as well as the median mass of raindrops $\tilde{m}$. In this way, evaporation of drops generally does not alter the type of the drop-size distribution, and the steady-state size distribution of drops in evaporating rains also obeys a lognormal law similar to that given by Eqs. (13) and (14) for mass- and number-weighted size distributions of raindrops respectively. Specifically, we found that Eqs. (13) and (14) can describe well the experimentally observed drop-size distribution even for evaporating rains, if the rain intensity is decreased from the actually measured value, as would be the case for a non-evaporating rain of a smaller intensity.

## 2. Comparison with experimental data

To compare the theory with experimental data, we analyzed experimental publications to determine what model of a disdrometer was used to measure the raindrop diameters, and how the raindrop size spectrograms and characteristic raindrop diameters were obtained. We processed the published time-averaged data and converted the experimental time-averaged spectrograms into experimental mass- and number-weighted size distributions and compared with our theoretical calculations.

For experimental data obtained by using the Joss–Waldvogel impact disdrometer [111] (JWD), the raindrop size spectrogram is constructed using 20 drop diameter bins (or classes) spanning between 0.36-5.37 mm for average bin values. The experimental number density of drops, $N(d_i)$, has units of 1/(m³·mm) and describes the class of drops $i$ with average diameter $d_i$, and is determined by the formula [88]:

$$N(d_i) = \frac{\Delta n_i}{A \Delta t U_i \Delta d_i}, \qquad (A21)$$

where $\Delta n_i$ is number of drops measured in drop size class $i$ during time interval $\Delta t$, $A = 0.005 \, m^2$ is the sampling area of the sensor, and $U_i$ is the fall velocity of drop with diameter $d_i$ obtained experimentally by Gunn and Kinzer [109] and $\Delta d_i$ is diameter interval of drop size class $i$ (bin width). To convert the number density of drops into a probability distribution function, we use the number rate of raindrops for a drop class $i$, $\Delta \dot{n}_i = \Delta n_i / \Delta t_i = N(d_i) A U_i \Delta d_i$, and fraction of drops $\Delta Q_i = \Delta \dot{n}_i / \sum \Delta \dot{n}_i$. From here, the number-weighted probability density for a drop class $i$, $q_{n,i}$, which has units of 1/mm, is given by:

$$q_{n,i}(d_i) = \frac{\Delta Q_i}{\Delta d_i} \qquad (A22)$$

The mass-weighted probability density for a drop class $i$, $q_{m,i}$, is determined by:

$$q_{m,i}(d_i) = q_{n,i}(d_i) \cdot \frac{d_i^3}{d_{\tilde{m}}^3}, \qquad (A23)$$

where the diameter of a drop with average mass, $d_{\bar{m}}$, is given by $d_{\bar{m}} = \tilde{d} \exp\left(-\frac{3}{2}\sigma_d^2\right)$ [64]. The experimental count and mass median diameters are determined from the respective expressions $\sum_i q_{n,i} \Delta d_i = 0.5$ and $\sum_i q_{m,i} \Delta d_i = 0.5$.

For experimental data obtained by using the PARSIVEL laser-optical disdrometer, the raindrop size spectrogram is constructed using 32 drop diameter bins spanning between 0.06-24.5 mm for average bin values. The experimental number density of drops, is determined by modified Eq. (A21) in which physical sampling area, $A$, is substituted for drop-size dependent effective sampling area [112], $A_i$. The physical sampling area of the PARSIVEL disdrometer [87] is a rectangle determined by the length $a = 180$ mm and the width $b = 30$ mm, so that $A = ab$, and the effective sampling area [113] is given by $A_i = (a - d_i)(b - d_i)$ for the drop class with diameter $d_i$.

The analysis of accuracy and measurement errors of the disdrometers is extensively reported in the literature, for example in ref. [56,82,83,113–117].

**REFERENCES**

[1] R. Graves, *The Greek Myths*, Complete and unabridged ed. in one volume, 1st ed. (Moyer Bell, Mt. Kisco, N.Y. :, 1988).
[2] Pliny the Elder, *Natural history*, Vol. IV (Harvard University Press, Cambridge, MA, 1963).
[3] Aristotle, *Meteorologica. With an English Translation by H.D.P. Lee.* (Harvard University Press, Cambridge, MA, 2014).
[4] H. H. Frisinger, *History of Meteorology: To 1800.* (American Meteorological Society, Boston, MA, 2018).
[5] G. B. P. R. Office, *Rerum Britannicarum Medii Ævi Scriptores, or, Chronicles and Memorials of Great Britain and Ireland during the Middle Ages* (1858).
[6] H. Schedel, Michael. Zellmann-Rohrer, C. T. Hadavas, and S. S. Nahas, *Liber chronicarum: Translation* (Selim S. Nahas Press, Boston, Massachusetts, 2010).
[7] P. Squatriti, *The Floods of 589 and Climate Change at the Beginning of the Middle Ages: An Italian Microhistory*, Speculum **85**, 799 (2010).
[8] A. Lawrence-Mathers, *Medieval Meteorology: Forecasting the Weather from Aristotle to the Almanac* (Cambridge University Press, 2020).
[9] J. Cunningham, *An Account of the Impressions and Casts of Drops of Rain, Discovered in the Quarries at Storeton Hill, Cheshire*, Proc. Geol. Soc. Lond. **3**, 99 (1839).
[10] W. Buckland, *On Recent and Fossil Semi-Circular Cavities Caused by Air Bubbles on the Surface of Soft Clay and Resembling Impressions of Rain-Drops*, Rep. Br. Assoc. Adv. Sci. Trans. Sect. 57 (1842).
[11] C. Lyell, *On Fossil Rain-Marks of the Recent, Triassic, and Carboniferous Periods*, Q. J. Geol. Soc. **7**, 238 (1851).
[12] J. C. Warren, *On Fossil Raindrops*, Proc. Boston Soc. Nat. Hist. **5**, 187 (1855).
[13] J. Wyman, *On the Formation of Rain Impressions in Clay*, Proc. Boston Soc. Nat. Hist. **5**, 253 (1855).
[14] W. S. Cassata and P. R. Renne, *Fossil Raindrops and Ancient Air*, Nature **484**, 322 (2012).
[15] D. Biello, *Primeval Precipitation*, Sci. Am. **306**, 26 (2012).
[16] D. Baird, *Revision of the Pennsylvanian and Permian Footprints Limnopus, Allopus and Baropus*, J. Paleontol. **26**, 832 (1952).
[17] S. M. Som, D. C. Catling, J. P. Harnmeijer, P. M. Polivka, and R. Buick, *Air Density 2.7 Billion Years Ago Limited to Less than Twice Modern Levels by Fossil Raindrop Imprints*, Nature **484**, 359 (2012).
[18] Z. Remin, T. Krogulec, T. Drela, and M. Surowski, *The Recognition of Hailstone Impressions in Clay-Rich Sediment: Experimental Results and Relation to the Neoproterozoic Case*, J. Sediment. Res. **84**, 543 (2014).
[19] R. Hueso and A. Sánchez-Lavega, *Methane Storms on Saturn's Moon Titan*, Nature **442**, 428 (2006).
[20] Nature Editor, *Methane Showers Leave Titan's Northern Reaches Gleaming*, Nature **565**, 539 (2019).
[21] R. D. Dhingra et al., *Observational Evidence for Summer Rainfall at Titan's North Pole*, Geophys. Res. Lett. **46**, 1205 (2019).
[22] R. A. Craddock and R. D. Lorenz, *The Changing Nature of Rainfall during the Early History of Mars*, Icarus **293**, 172 (2017).
[23] G. Stucky de Quay, T. A. Goudge, and C. I. Fassett, *Precipitation and Aridity Constraints from Paleolakes on Early Mars*, Geology **48**, 1189 (2020).
[24] R. A. Craddock and A. D. Howard, *The Case for Rainfall on a Warm, Wet Early Mars*, J. Geophys. Res. Planets **107**, 21 (2002).
[25] P. Gao, X. Zhang, D. Crisp, C. G. Bardeen, and Y. L. Yung, *Bimodal Distribution of Sulfuric Acid Aerosols in the Upper Haze of Venus*, Icarus **231**, 83 (2014).
[26] H. F. Wilson and B. Militzer, *Sequestration of Noble Gases in Giant Planet Interiors*, Phys. Rev. Lett. **104**, 121101 (2010).

[27] M. Ross, *The Ice Layer in Uranus and Neptune - Diamonds in the Sky?*, Nature **292**, 435 (1981).
[28] S. Frydrych et al., *Demonstration of X-Ray Thomson Scattering as Diagnostics for Miscibility in Warm Dense Matter*, Nat. Commun. **11**, 2620 (2020).
[29] D. Ehrenreich et al., *Nightside Condensation of Iron in an Ultrahot Giant Exoplanet*, Nature **580**, 597 (2020).
[30] S. Battersby, *A Planet so Hot It Rains Iron*, New Sci. **177**, 20 (2003).
[31] R. Garner, *NASA Hubble finds a true blue planet*, http://www.nasa.gov/content/nasa-hubble-finds-a-true-blue-planet.
[32] J. Wenz, *Gemstone Clouds*, Astronomy **45**, 12 (2017).
[33] B. Benneke et al., *Water Vapor and Clouds on the Habitable-Zone Sub-Neptune Exoplanet K2-18b*, Astrophys. J. Lett. **887**, L14 (2019).
[34] W. A. Bentley, *Studies of Raindrops and Raindrop Phenomena*, Mon. Weather Rev. **32**, 450 (1904).
[35] I. Langmuir, *The Production of Rain by a Chain Reaction in Cumulus Clouds at Temperatures above Freezing*, J. Atmospheric Sci. **5**, 175 (1948).
[36] K. Loftus and R. D. Wordsworth, *The Physics of Falling Raindrops in Diverse Planetary Atmospheres*, J. Geophys. Res. Planets **126**, e2020JE006653 (2021).
[37] H. R. Pruppacher and J. D. Klett, *Microphysics of Clouds and Precipitation*, Vol. 18 (Springer Netherlands, Dordrecht, 2010).
[38] D. C. Blanchard, *The Behavior of Water Drops at Terminal Velocity in Air*, Trans. Am. Geophys. Union **31**, 836 (1950).
[39] E. M. F. D'Albe and M. S. Hidayetulla, *The Break-up of Large Water Drops Falling at Terminal Velocity in Free Air*, Q. J. R. Meteorol. Soc. **81**, 610 (1955).
[40] M. Komabayasi, T. Gonda, and K. Isono, *Life Time of Water Drops before Breaking and Size Distribution of Fragment Droplets*, J. Meteorol. Soc. Jpn. Ser II **42**, 330 (1964).
[41] D. C. Blanchard and A. T. Spencer, *Experiments on the Generation of Raindrop-Size Distributions by Drop Breakup*, J. Atmospheric Sci. **27**, 101 (1970).
[42] M. Komabayasi, *Probability of Disintegration of Water Drops as a Factor Determining Size Distribution of Raindrops*, in *Proceedings of International Conference on Cloud Physics* (Tokyo and Sapporo, 1965), pp. 260–264.
[43] E. Villermaux, *Fragmentation*, Annu. Rev. Fluid Mech. **39**, 419 (2007).
[44] E. Villermaux, *Fragmentation versus Cohesion*, J. Fluid Mech. **898**, P1 (2020).
[45] E. Villermaux and B. Bossa, *Single-Drop Fragmentation Determines Size Distribution of Raindrops*, Nat. Phys. **5**, 697 (2009).
[46] E. Villermaux and B. Bossa, *Size Distribution of Raindrops*, Nat. Phys. **6**, 4 (2010).
[47] R. D. Lorenz, *The Life, Death and Afterlife of a Raindrop on Titan*, Planet. Space Sci. **41**, 647 (1993).
[48] A. M. Palumbo, J. W. Head, and L. Wilson, *Rainfall on Noachian Mars: Nature, Timing, and Influence on Geologic Processes and Climate History*, Icarus **347**, 113782 (2020).
[49] A. P. Barros, O. P. Prat, and F. Y. Testik, *Size Distribution of Raindrops*, Nat. Phys. **6**, 232 (2010).
[50] G. M. McFarquhar, *Raindrop Size Distribution and Evolution*, in *Rainfall: State of the Science*, edited by F. Y. Testik and M. Gebremichael, Vol. 191 (2010), pp. 49–60.
[51] F. Y. Testik and M. K. Rahman, *First in Situ Observations of Binary Raindrop Collisions*, Geophys. Res. Lett. **44**, 1175 (2017).
[52] F. Y. Testik and B. Pei, *Wind Effects on the Shape of Raindrop Size Distribution*, J. Hydrometeorol. **18**, 1285 (2017).
[53] R. List and G. M. McFarquhar, *The Role of Breakup and Coalescence in the Three-Peak Equilibrium Distribution of Raindrops*, J. Atmospheric Sci. **47**, 2274 (1990).

[54] E. Adirosi, L. Baldini, and A. Tokay, *Rainfall and DSD Parameters Comparison between Micro Rain Radar, Two-Dimensional Video and Parsivel2 Disdrometers, and S-Band Dual-Polarization Radar*, J. Atmospheric Ocean. Technol. **37**, 621 (2020).

[55] A. Tokay, L. P. D'Adderio, D. A. Marks, J. L. Pippitt, D. B. Wolff, and W. A. Petersen, *Comparison of Raindrop Size Distribution between NASA's s-Band Polarimetric Radar and Two-Dimensional Video Disdrometers*, J. Appl. Meteorol. Climatol. **59**, 517 (2020).

[56] W.-Y. Chang, G. Lee, B. J.-D. Jou, W.-C. Lee, P.-L. Lin, and C.-K. Yu, *Uncertainty in Measured Raindrop Size Distributions from Four Types of Collocated Instruments*, Remote Sens. **12**, 1167 (2020).

[57] B. J. Miriovsky et al., *An Experimental Study of Small-Scale Variability of Radar Reflectivity Using Disdrometer Observations*, J. Appl. Meteorol. Climatol. **43**, 106 (2004).

[58] M. Steiner and J. A. Smith, *Scale Dependence of Radar-Rainfall Rates—An Assessment Based on Raindrop Spectra*, J. Hydrometeorol. **5**, 1171 (2004).

[59] A. Tokay, L. P. D'Adderio, D. B. Wolff, and W. A. Petersen, *Development and Evaluation of the Raindrop Size Distribution Parameters for the NASA Global Precipitation Measurement Mission Ground Validation Program*, J. Atmospheric Ocean. Technol. **37**, 115 (2020).

[60] C. R. Williams et al., *Describing the Shape of Raindrop Size Distributions Using Uncorrelated Raindrop Mass Spectrum Parameters*, J. Appl. Meteorol. Climatol. **53**, 1282 (2014).

[61] M. Mezhericher and H. A. Stone, *Possible, Impossible, and Expected Diameters and Production Rates of Droplets in Aerosols and Sprays*, Phys. Rev. Fluids **7**, 063602 (2022).

[62] C. Borgnakke and R. E. Sonntag, *Fundamentals of Thermodynamics*, 10th ed. (John Wiley & Sons Inc., USA, 2020).

[63] R. B. Bird, W. E. Stewart, and E. N. Lightfoot, *Transport Phenomena* (Wiley, 2006).

[64] W. C. Hinds, *Aerosol Technology: Properties, Behavior, and Measurement of Airborne Particles* (Wiley, 2012).

[65] T. Hatch and S. P. Choate, *Statistical Description of the Size Properties of Non Uniform Particulate Substances*, J. Frankl. Inst. **207**, 369 (1929).

[66] B. K. Jones, J. R. Saylor, and F. Y. Testik, *Raindrop Morphodynamics*, in *Rainfall: State of the Science*, edited by F. Y. Testik and M. Gebremichael, Vol. 191 (2010), pp. 7–28.

[67] Van den Broeck C., *Stochastic Thermodynamics: A Brief Introduction*, Proc. Int. Sch. Phys. Enrico Fermi **184**, 155 (2013).

[68] L. D. Landau and E. M. Lifshitz, *Statistical Physics: Volume 5* (Elsevier Science, 2013).

[69] K. V. Beard, H. T. Ochs, and R. J. Kubesh, *Natural Oscillations of Small Raindrops*, Nature **342**, 408 (1989).

[70] H. R. Pruppacher and K. V. Beard, *A wind tunnel investigation of the internal circulation and shape of water drops falling at terminal velocity in air*, Q. J. R. Meteorol. Soc. **96**, 247 (1970).

[71] J. R. Saylor and B. K. Jones, *The Existence of Vortices in the Wakes of Simulated Raindrops*, Phys. Fluids **17**, 031706 (2005).

[72] M. Thurai, V. Bringi, P. Gatlin, and M. Wingo, *Raindrop Fall Velocity in Turbulent Flow: An Observational Study*, Adv. Sci. Res. **18**, 33 (2021).

[73] J. S. Marshall and W. M. K. Palmer, *The Distribution of Raindrops with Size*, J. Meteorol. **5**, 165 (1948).

[74] J. O. Laws and D. A. Parsons, *The Relation of Raindrop-Size to Intensity*, Trans. Am. Geophys. Union **24**, 452 (1943).

[75] G. Feingold and Z. Levin, *The Lognormal Fit to Raindrop Spectra from Frontal Convective Clouds in Israel*, J. Appl. Meteorol. Climatol. **25**, 1346 (1986).

[76] H. Sauvageot and J.-P. Lacaux, *The Shape of Averaged Drop Size Distributions*, J. Atmospheric Sci. **52**, 1070 (1995).

[77] G. E. Uhlenbeck and L. S. Ornstein, *On the Theory of the Brownian Motion*, Phys. Rev. **36**, 823 (1930).

[78] H. Risken, *The Fokker-Planck Equation: Methods of Solution and Applications*, 2nd ed (Springer-Verlag, New York, 1996).
[79] J. A. Smith and R. D. D. Veaux, *A Stochastic Model Relating Rainfall Intensity to Raindrop Processes*, Water Resour. Res. **30**, 651 (1994).
[80] E. Baltas, D. Panagos, and M. Mimikou, *Statistical Analysis of the Raindrop Size Distribution Using Disdrometer Data*, Hydrology **3**, 1 (2016).
[81] F. Pukelsheim, *The Three Sigma Rule*, Am. Stat. **48**, 88 (1994).
[82] J. L. Jaffrain and A. Berne, *Experimental Quantification of the Sampling Uncertainty Associated with Measurements from PARSIVEL Disdrometers*, J. Hydrometeorol. **12**, 19 (2011).
[83] J. Joss and A. Waldvogel, *Raindrop Size Distribution and Sampling Size Errors*, J. Atmospheric Sci. **26**, 566 (1969).
[84] A. Tokay, A. Kruger, W. F. Krajewski, P. A. Kucera, and A. J. P. Filho, *Measurements of Drop Size Distribution in the Southwestern Amazon Basin*, J. Geophys. Res. Atmospheres **107**, LBA 19 (2002).
[85] S. Niu, X. Jia, J. Sang, X. Liu, C. Lu, and Y. Liu, *Distributions of Raindrop Sizes and Fall Velocities in a Semiarid Plateau Climate: Convective versus Stratiform Rains*, J. Appl. Meteorol. Climatol. **49**, 632 (2010).
[86] World Meteorological Organization, *Guide to Hydrological Practices*, WMO-No. 168 (1994).
[87] Operating Instructions: Present Weather Sensor OTT Parsivel2, OTT HydroMet, 2016.
[88] JWD RD-80 Disdrometer. Disdrodata - User Guide 4.0.Pdf, DISTROMET LTD, www.distromet.com, 2018.
[89] P. N. Gatlin, M. Thurai, V. N. Bringi, W. Petersen, D. Wolff, A. Tokay, L. Carey, and M. Wingo, *Searching for Large Raindrops: A Global Summary of Two-Dimensional Video Disdrometer Observations*, J. Appl. Meteorol. Climatol. **54**, 1069 (2015).
[90] Y. Fujiyoshi, I. Yamamura, N. Nagumo, K. Nakagawa, K. Muramoto, and T. Shimomai, *The Maximum Size of Raindrops – Can It Be a Proxy of Precipitation Climatology?*, in (Cancun, Mexico, 2008).
[91] P. V. Hobbs and A. L. Rangno, *Super-Large Raindrops*, Geophys. Res. Lett. **31**, (2004).
[92] K. V. Beard, D. B. Johnson, and D. Baumgardner, *Aircraft Observations of Large Raindrops in Warm, Shallow, Convective Clouds*, Geophys. Res. Lett. **13**, 991 (1986).
[93] R. Clift, J. R. Grace, and M. E. Weber, *Bubbles, Drops, and Particles* (Academic Press, New York, 1978).
[94] D. Chicco, M. J. Warrens, and G. Jurman, *The Coefficient of Determination R-Squared Is More Informative than SMAPE, MAE, MAPE, MSE and RMSE in Regression Analysis Evaluation*, PeerJ Comput. Sci. **7**, e623 (2021).
[95] C. W. Ulbrich, *The Effects of Drop Size Distribution Truncation on Rainfall Integral Parameters and Empirical Relations*, J. Appl. Meteorol. Climatol. **24**, 580 (1985).
[96] A. D. Ochou, A. Nzeukou, and H. Sauvageot, *Parametrization of Drop Size Distribution with Rain Rate*, Atmospheric Res. **84**, 58 (2007).
[97] S.-H. Suh, C.-H. You, and D.-I. Lee, *Climatological Characteristics of Raindrop Size Distributions in Busan, Republic of Korea*, Hydrol. Earth Syst. Sci. **20**, 193 (2016).
[98] F. Y. Testik and A. P. Barros, *Toward Elucidating the Microstructure of Warm Rainfall: A Survey*, Rev. Geophys. **45**, (2007).
[99] E. Villermaux and F. Eloi, *The Distribution of Raindrops Speeds*, Geophys. Res. Lett. **38**, (2011).
[100] R. A. Houze, P. V. Hobbs, P. H. Herzegh, and D. B. Parsons, *Size Distributions of Precipitation Particles in Frontal Clouds*, J. Atmospheric Sci. **36**, 156 (1979).
[101] D. A. de Wolf, *On the Laws-Parsons Distribution of Raindrop Sizes*, Radio Sci. **36**, 639 (2001).
[102] A. Waldvogel, *The N0 Jump of Raindrop Spectra*, J. Atmospheric Sci. **31**, 1067 (1974).
[103] C. W. Ulbrich, *Natural Variations in the Analytical Form of the Raindrop Size Distribution*, J. Clim. Appl. Meteorol. **22**, 1764 (1983).

[104] A. H. Lefebvre and V. G. McDonell, *Atomization and Sprays*, Second edition (CRC Press, Boca Raton, 2017).
[105] R. List, C. F. MacNeil, and J. D. McTaggart-Cowan, *Laboratory Investigations of Temporary Collisions of Raindrops*, J. Geophys. Res. 1896-1977 **75**, 7573 (1970).
[106] T. B. Low and R. List, *Collision, Coalescence and Breakup of Raindrops. Part I: Experimentally Established Coalescence Efficiencies and Fragment Size Distributions in Breakup*, J. Atmospheric Sci. **39**, 1591 (1982).
[107] R. R. Rogers, *Raindrop Collision Rates*, J. Atmospheric Sci. **46**, 2469 (1989).
[108] G. M. McFarquhar and R. List, *The Raindrop Mean Free Path and Collision Rate Dependence on Rainrate for Three-Peak Equilibrium and Marshall–Palmer Distributions*, J. Atmospheric Sci. **48**, 1999 (1991).
[109] R. Gunn and G. D. Kinzer, *The Terminal Velocity of Fall for Water Droplets in Stagnant Air*, J. Atmospheric Sci. **6**, 243 (1949).
[110] E. W. Grafarend, *Linear and Nonlinear Models: Fixed Effects, Random Effects, and Mixed Models* (Walter de Gruyter, 2006).
[111] J. Joss and A. Waldvogel, *Ein Spektrograph für Niederschlagstropfen mit automatischer Auswertung*, Pure Appl. Geophys. **68**, 240 (1967).
[112] T. H. Raupach and A. Berne, *Correction of Raindrop Size Distributions Measured by Parsivel Disdrometers, Using a Two-Dimensional Video Disdrometer as a Reference*, Atmospheric Meas. Tech. **8**, 343 (2015).
[113] R. Fraile, A. Castro, M. Fernández-Raga, C. Palencia, and A. I. Calvo, *Error in the Sampling Area of an Optical Disdrometer: Consequences in Computing Rain Variables*, https://doi.org/10.1155/2013/369450.
[114] H. S. Gertzman and D. Atlas, *Sampling Errors in the Measurement of Rain and Hail Parameters*, J. Geophys. Res. 1896-1977 **82**, 4955 (1977).
[115] A. Tokay, K. R. Wolff, P. Bashor, and O. K. Dursun, *On the Measurement Errors of the Joss-Waldvogel Disdrometer.*, in (NTRS - NASA Technical Reports Server, https://ntrs.nasa.gov/citations/20030053178, 2003), p. 4.
[116] A. Tokay, P. G. Bashor, and K. R. Wolff, *Error Characteristics of Rainfall Measurements by Collocated Joss–Waldvogel Disdrometers*, J. Atmospheric Ocean. Technol. **22**, 513 (2005).
[117] S. L. M. Neto, Y. Sakagami, I. C. da Costa, E. B. Pereira, J. C. Thomaz Junior, and A. Von Wangenheim, Uncertainty Analysis of Disdrometer Model PARSIVEL2 for Rainfall Amount, Joint Research Report of INPECCST, INPE-CPTEC, UFSCEMC- LEPTEN/LABSOLAR, IFSC-Florianópolis and UFSCINE- LAPIX/INCOD No. o sid.inpe.br/mtc-m21b/2016/02.12.11.38-RPQ, INPE, 2015.

**FIGURES**

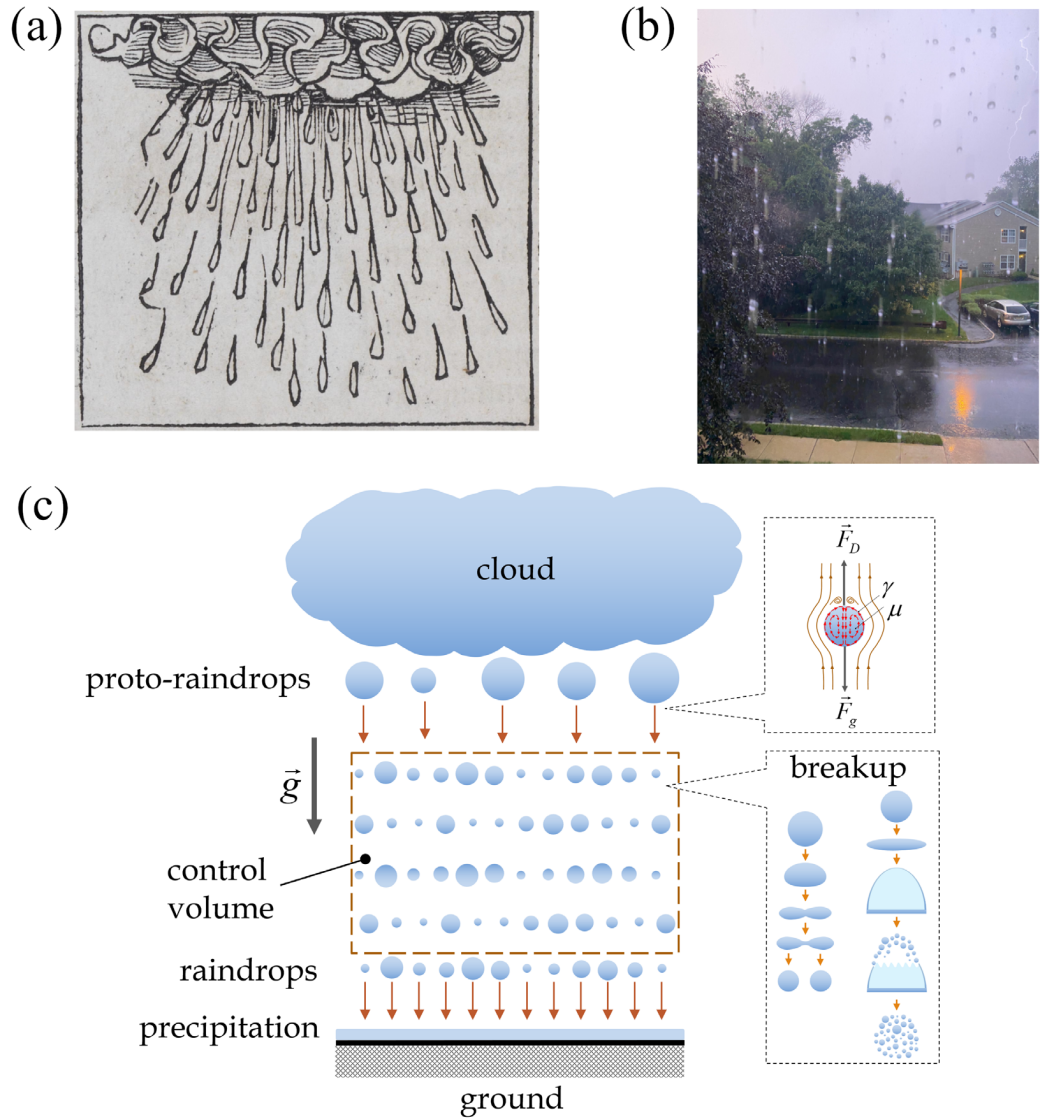

FIG. 1. (a) Illustration of a heavy rain that resulted in an extensive flood in Italy circa 587 C.E., depicted as the woodcut in the Nuremberg Chronicle, 1493 [6]. (b) Different size of drops in a rain photographed in Princeton, NJ, USA on July 6, 2021. The falling drops were "frozen" owing to a stroboscopic backlight effect produced by the lightning flash visible in the top right corner of the image. (c) Schematic diagram of rainfall for the theory development. Initially formed proto-raindrops upon falling from a cloud are subjected to air drag and Earth's gravity. These forces can result in drop breakup either into two or more daughter drops, depending on the breakup mechanism (e.g., illustrated here are oscillatory and bag breakups, and collisional and other types of breakup are not shown [104]). The red dashed line outlines the boundary of the control volume for thermodynamic analysis, $\vec{g}$ is gravitational acceleration, $\vec{F}_g$ and $\vec{F}_D$ are gravity and drag forces acting on a falling raindrop, and $\gamma$ and $\mu$ are surface tension and dynamic viscosity of the raindrop, respectively.

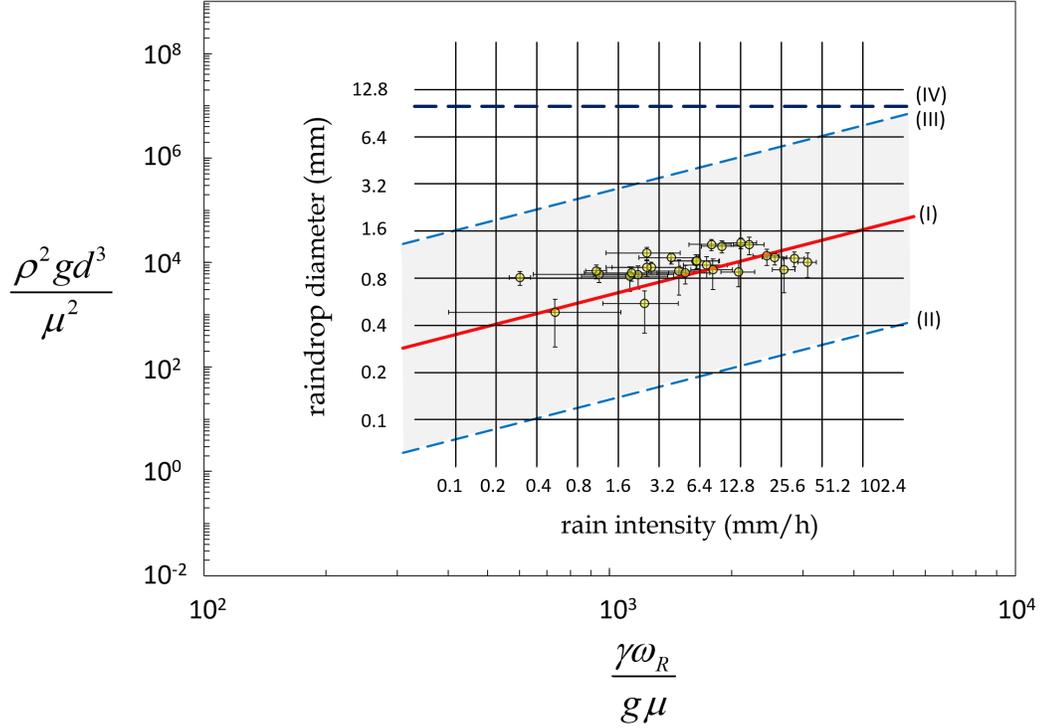

FIG. 2. Theoretical diagram of expected diameters of raindrops as a function of rain intensity. The theoretical count median diameters, $\tilde{d}_n$, are given by the line (I) following the Eq. (8), where the standard deviation, $\sigma_d$, of raindrop size distribution was assumed to be constant and set to the geometric mean value $\sigma_d = \frac{1}{3\sqrt{2}}\ln 10$ for all rain intensities. The shaded region between the lines (II) and (III) denotes the range of possible drop diameters for a rain of given intensity. The line (II) describes possible minimum diameters of raindrops, $d_{min}$, and is given by $\frac{\rho^2 g d_{min}^3}{\mu^2} = 0.01 \frac{\rho^2 g \tilde{d}_n^3}{\mu^2}$, which leads to $d_{min} \approx 0.22 \tilde{d}_n$. The line (III) describes possible maximum diameters of raindrops, $d_{max}$, and is given by $\frac{\rho^2 g d_{max}^3}{\mu^2} = 100 \frac{\rho^2 g \tilde{d}_n^3}{\mu^2}$, which leads to $d_{max} \approx 4.64 \tilde{d}_n$. The line (IV) corresponds to the largest ever observed raindrop diameter of 9.7 mm [89]. The circles correspond to the experimental count median diameters, which were obtained from the analysis of the published time-averaged spectrograms of rains with intensities 0.4-40 mm/h recorded by disdrometers in different parts of the world (the respective reports are in the literature [80,84,85] and in additional references that are summarized in the Table S1 of the Supplementary Information). The error bars represent measurement uncertainties that we evaluated based on the disdrometer model and measurement error analysis [82,83]. The theoretical connection between the rain intensity and the rain frequency, $\omega_R$, is given by Eq. (9). Below the line (II), there is insufficient amount of available energy that can be supplied by the aerodynamic drag and gravity forces to compete with the rainwater surface tension and viscosity forces, so that raindrops smaller than $d_{min}$ cannot be formed. Above the line (III) there is substantial excess of available energy from aerodynamic drag and gravity forces over the surface tension and viscosity of rainwater, so that every drop larger than $d_{max}$ breaks apart into smaller drops. The diagram is based on the same principles of possible, impossible and expected drop diameters which were described recently for similar diagrams of droplet diameters of aerosols and sprays [61].

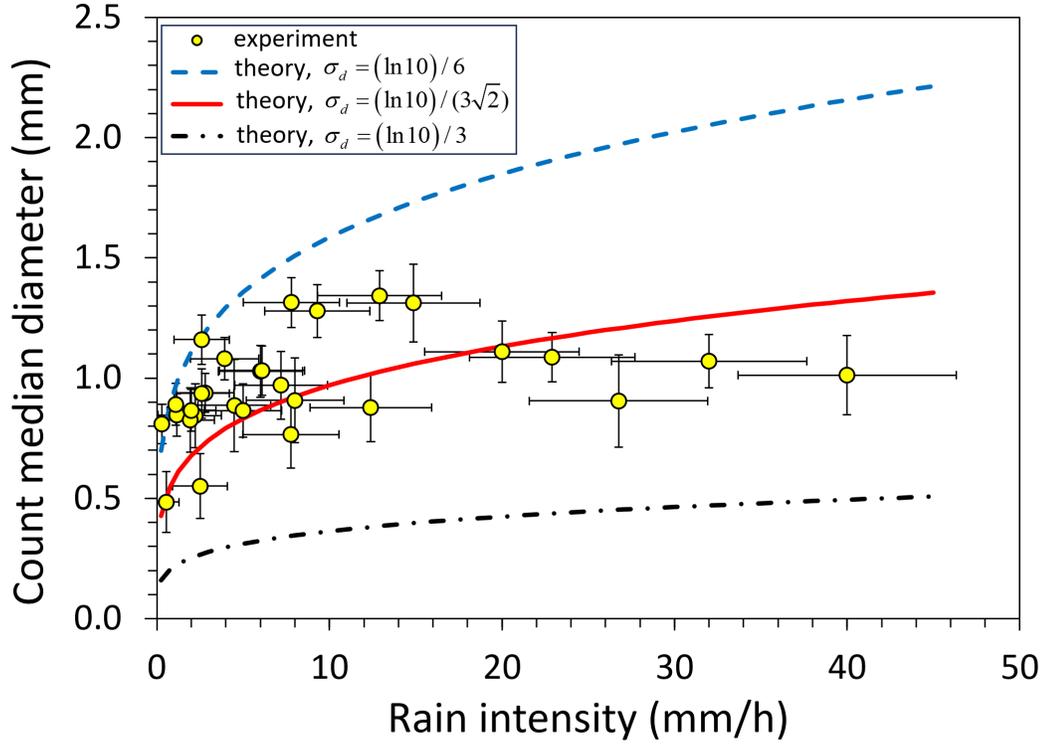

FIG. 3. Count median diameters of raindrops at various rain intensities. The circles correspond to experimental data from time-averaged rain spectrograms for rains precipitated at different geographic locations around the world (these date are published in the literature [80,84,85] and in additional references given in the Supplementary Information). Error bars represent measurement uncertainties that we evaluated based on the disdrometer model and error analysis reported in the literature [82,83]. The lines were calculated by Eq. (12) and the Hatch-Choate relationship [64,65] $\tilde{d}_n = \tilde{d}\exp(-3\sigma_d^2)$, and applying three different standard deviations $\sigma_d = \frac{1}{6}\ln 10$, $\sigma_d = \frac{1}{3\sqrt{2}}\ln 10$, and $\sigma_d = \frac{1}{3}\ln 10$.

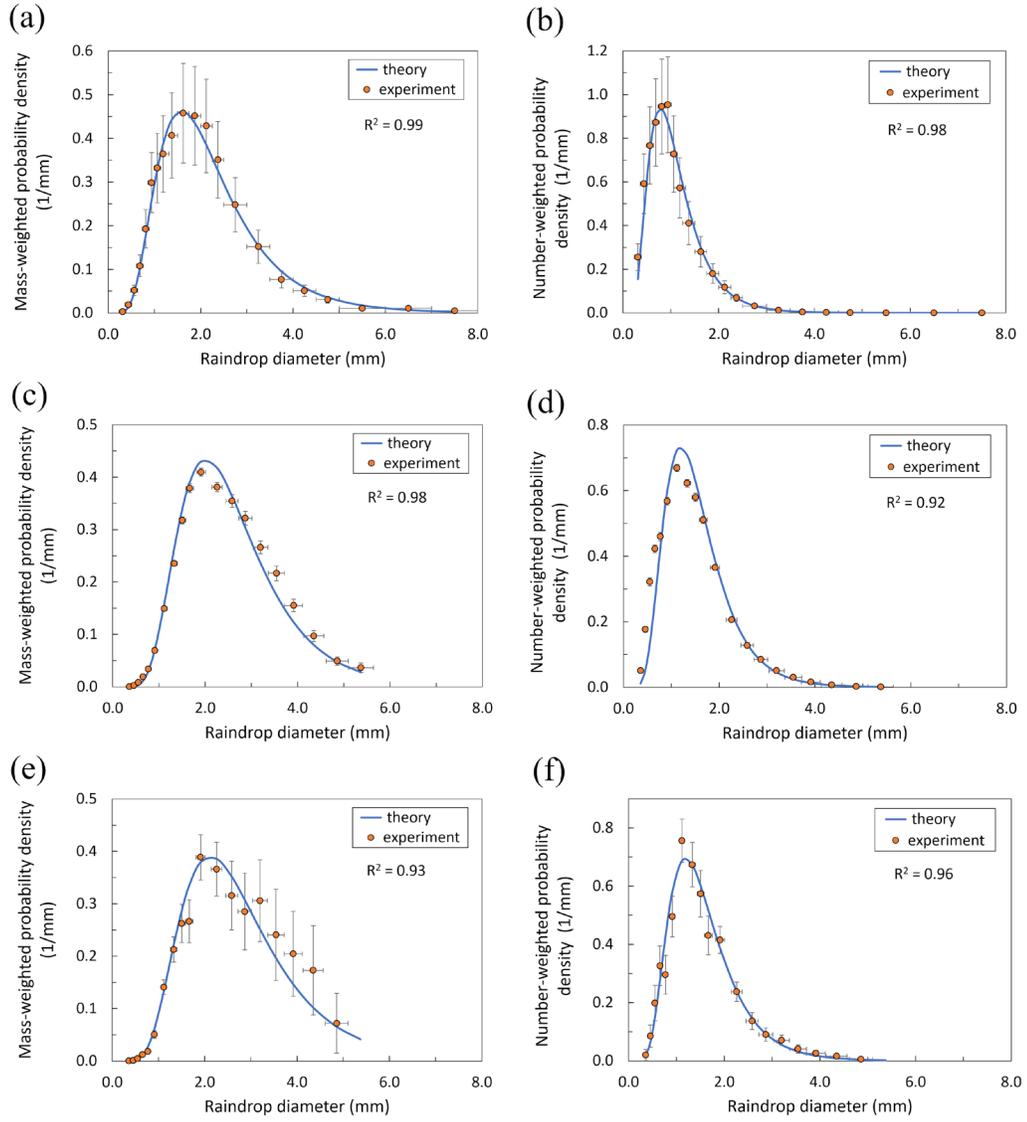

FIG. 4. Mass-weighted and number-weighted size distributions of raindrops for exemplary rains. (a), (b) Rain in China with average intensity 4.5 mm/h [85]; (c), (d) rain in Greece with average intensity 9.3 mm/h [80]; (e), (f) rain in the Amazon basin with average intensity 12.9 mm/h [84]. The circles indicate experimental time-averaged data points obtained in the studies of other researchers by averaging multiple instantaneously observed raindrop spectrograms, which were sampled by disdrometers with fixed time intervals during a rainfall event. The error bars reflect the measurement uncertainties, which we calculated according to the error analysis reported in the literature [82,83]. The theoretical lines of the mass-weighted probability density functions were calculated according to Eqs. (12) and (13). The lines of the number-weighted probability density function were found from Eqs. (12) and (14), and applying the Hatch-Choate relationship [64] $\tilde{d}_n = \tilde{d} \exp(-3\sigma_d^2)$. The values of the standard deviation of logarithm of raindrop diameter, $\sigma_d$, were selected from the theoretical range $\frac{1}{6}\ln 10 \leq \sigma_d \leq \frac{1}{3}\ln 10$, and were set to $\sigma_d = 0.49$ (panels a, b), $\sigma_d = 0.42$ (c, d) and $\sigma_d = 0.44$ (e, f). The coefficient of determination [94], $R^2$, characterizes the agreement between the calculated and experimental points.

# Supplementary Information

# Stochastic Theory of the Size Distribution of Raindrops

Maksim Mezhericher* and Howard A. Stone

Department of Mechanical and Aerospace Engineering, Princeton University
Princeton, NJ 08544, USA

* Corresponding author, email: maksymm@princeton.edu

**This file includes:**

- Supplementary text.
- Table S1, summarizing the measured count median diameters of raindrops.
- Figures S1 to S13, demonstrating the overlap between the theoretical and experimental probability density functions of drop-size distributions for different rains.
- Figure S14, the diagram of mass median diameters of raindrops, demonstrating the overlap between the theory and experimental data.
- SI References.

## 1. Equations of Size Distribution of Raindrops

1) The mass-weighted probability density function of a logarithm of a normalized raindrop diameter is given by a normal Gauss-Laplace distribution function:

$$q_m(\ln \delta) = \frac{1}{\sigma_d \sqrt{2\pi}} \exp\left[-\frac{1}{2}\left(\frac{\ln \delta}{\sigma_d}\right)^2\right], \quad (S1)$$

where $\delta = d/\tilde{d}$ is mass-weighted normalized raindrop diameter, $d$ is raindrop diameter, and $\tilde{d}$ is mass median diameter (so called "MMD" in aerosols and sprays [1]) of a raindrop. The mass median raindrop diameter is connected to rain intensity by the following expression:

$$\tilde{d} = \left[\frac{9\gamma^2}{\pi^2 \rho^2 g^3} \exp\left(-3\sigma_d^2 - 2\right)\right]^{1/9} \phi^{2/9}, \quad (S2)$$

where $\phi = I\hat{A}$ is volumetric flow rate of rain precipitating on a unit square area of the ground. By the definition $\phi$ is an intensive parameter, as it is obtained from rain intensity (also called "rain rate"), $I$, and unit square area, $\hat{A}$.

2) The mass-weighted probability density function of a normalized raindrop diameter is given by a lognormal distribution function:

$$q_m(\delta) = \frac{1}{\delta \sigma_d \sqrt{2\pi}} \exp\left[-\frac{1}{2}\left(\frac{\ln \delta}{\sigma_d}\right)^2\right] \quad (S3)$$

3) The mass-weighted probability density function of a raindrop diameter is given by a lognormal distribution function:

$$q_m(d) = \frac{1}{d \sigma_d \sqrt{2\pi}} \exp\left[-\frac{1}{2}\left(\frac{\ln d/\tilde{d}}{\sigma_d}\right)^2\right] \quad (S4)$$

4) The number-weighted probability density function of a logarithm of a normalized raindrop diameter is given by a normal Gauss-Laplace distribution function:

$$q_n(\ln \delta_n) = \frac{1}{\sigma_d \sqrt{2\pi}} \exp\left[-\frac{1}{2}\left(\frac{\ln \delta_n}{\sigma_d}\right)^2\right], \quad (S5)$$

where $\delta_n = d/\tilde{d}_n$ is number-weighted normalized raindrop diameter, $d$ is raindrop diameter, and $\tilde{d}_n$ is count median diameter (so called "CMD" in aerosols and sprays, also equal to geometric mean diameter [1]) of raindrop size distribution. The count median and mass median diameters of raindrops for a lognormal size distribution are connected by Hatch-Choate equation [1,2]:

$$\tilde{d}_n = \tilde{d} \exp\left(-3\sigma_d^2\right) \quad (S6)$$

Combining Eq. (S2) and (S6), we get:

$$\tilde{d}_n = \left[\frac{9\gamma^2}{\pi^2 \rho^2 g^3} \exp\left(-30\sigma_d^2 - 2\right)\right]^{1/9} \phi^{2/9} \quad (S7)$$

5) The number-weighted probability density function of a normalized raindrop diameter is given by a lognormal distribution function:

$$q_n(\delta_n) = \frac{1}{\delta_n \sigma_d \sqrt{2\pi}} \exp\left[-\frac{1}{2}\left(\frac{\ln \delta_n}{\sigma_d}\right)^2\right] \tag{S8}$$

6) The number-weighted probability density function of a raindrop diameter is given by a lognormal distribution function:

$$q_n(d) = \frac{1}{d \sigma_d \sqrt{2\pi}} \exp\left[-\frac{1}{2}\left(\frac{\ln d/\tilde{d}_n}{\sigma_d}\right)^2\right] \tag{S9}$$

7) The standard deviation $\sigma_d$, which is the measure of the magnitude of stochastic energy fluctuations in a rain, is theoretically predicted to be in the range $\frac{1}{6}\ln 10 \leq \sigma_d \leq \frac{1}{3}\ln 10$ ($0.38 \leq \sigma_d \leq 0.77$), as described in the main article text. For each rain, the value of standards deviation was chosen from the above range to conform the theoretical curves with the experimental data.

## 2. Comparison Between the Theory and Experiments

### 2.1. Measured Count Median Diameters of Raindrops

**TABLE S1**. Summary of the measured count median diameters (CMD) of raindrops, which were utilized for the theory validation. The CMD values were found by processing the published rain spectrograms using the procedure described in Appendix of the main text.

| # | Literature reference | Time-averaged rain intensity, mm/h | Uncertainty of time-averaged rain intensity, mm/h | CMD, mm | Uncertainty of CMD, mm |
|---|---|---|---|---|---|
| 1 | Niu et al. [3] | 4.49 | 2.12 | 0.89 | 0.19 |
| 2 | Baltas et al. [4] | 9.30 | 3.05 | 1.28 | 0.11 |
| 3 | Baltas et al. [4] | 2.80 | 1.67 | 0.94 | 0.08 |
| 4 | Tokay et al. [5] | 7.20 | 2.68 | 0.97 | 0.14 |
| 5 | Tokay et al. [6] | 12.90 | 3.59 | 1.34 | 0.10 |
| 6 | Tokay et al. [6] | 1.15 | 1.07 | 0.85 | 0.09 |
| 7 | Tokay et al. [7] | 7.80 | 2.79 | 1.31 | 0.10 |
| 8 | Loh et al. [8] | 26.76 | 5.17 | 0.90 | 0.19 |
| 9 | Suh et al. [9] | 8.00 | 2.83 | 0.91 | 0.18 |
| 10 | Smith [10] | 2.24 | 1.50 | 0.84 | 0.13 |
| 11 | Tokay et al. [7] | 1.10 | 0.20 | 0.89 | 0.09 |
| 12 | D'Adderio et al. [11] | 40.00 | 6.32 | 1.01 | 0.16 |
| 13 | Niu et al. [3] | 0.54 | 0.74 | 0.48 | 0.13 |
| 14 | Tokay et al. [6] | 32.00 | 5.66 | 1.07 | 0.11 |
| 15 | Tokay et al. [7] | 0.30 | 0.05 | 0.81 | 0.08 |
| 16 | Tokay et al. [5] | 12.40 | 3.52 | 0.88 | 0.14 |
| 17 | Suh et al. [9] | 2.50 | 1.58 | 0.55 | 0.13 |
| 18 | Loh et al. [8] | 1.94 | 1.39 | 0.83 | 0.13 |
| 19 | Zeng et al. [12] | 14.87 | 3.86 | 1.31 | 0.16 |
| 20 | Lam et al. [13] | 22.91 | 4.79 | 1.09 | 0.10 |
| 21 | Cha and Yum [14] | 5.00 | 2.24 | 0.87 | 0.11 |
| 22 | Tokay et al. [6] | 2.60 | 1.61 | 1.16 | 0.10 |
| 23 | Tokay et al. [6] | 6.00 | 2.45 | 1.03 | 0.11 |
| 24 | Tokay et al. [7] | 3.93 | 1.98 | 1.08 | 0.09 |
| 25 | Tokay et al. [7] | 2.00 | 1.41 | 0.87 | 0.09 |
| 26 | D'Adderio et al. [11] | 20.00 | 4.47 | 1.11 | 0.13 |
| 27 | Tokay et al. [7] | 6.10 | 2.47 | 1.03 | 0.10 |
| 28 | Tokay et al. [7] | 2.60 | 1.61 | 0.94 | 0.10 |
| 29 | Zeng et al. [12] | 7.77 | 2.79 | 0.77 | 0.14 |

To evaluate the measurement uncertainties of the time-averaged rain intensities, the uncertainty, $\Delta I$, was assumed to follow the square root scaling with the rain intensity, $I$, i.e. $\Delta I \propto \sqrt{I}$. In this work, for rain intensities with units of mm/h, the proportionality coefficient was set to $1 \text{ mm}^{0.5}/\text{h}^{0.5}$, to obtain the values of measurement uncertainties of time-averaged rain intensities given in Table S1. For the measurement uncertainties of count median diameters (CMD) of raindrops, the propagation of uncertainty was calculated using the general method involving a total differential of a function of independent variables [15], which was applied to the expression $\sum_i q_{n,i} \Delta d_i = 0.5$ and Eqs. (A21) and (A22) determining the measured CMD (see Appendix of the main text).

## 2.2. Size Distribution of Raindrops

For better readability, on the below figures the experimental points are shown without the error bars which are typically used to reflect the measurement uncertainties. The respective analysis of the measurement errors of the disdrometers is extensively reported in the literature, for example [16–23].

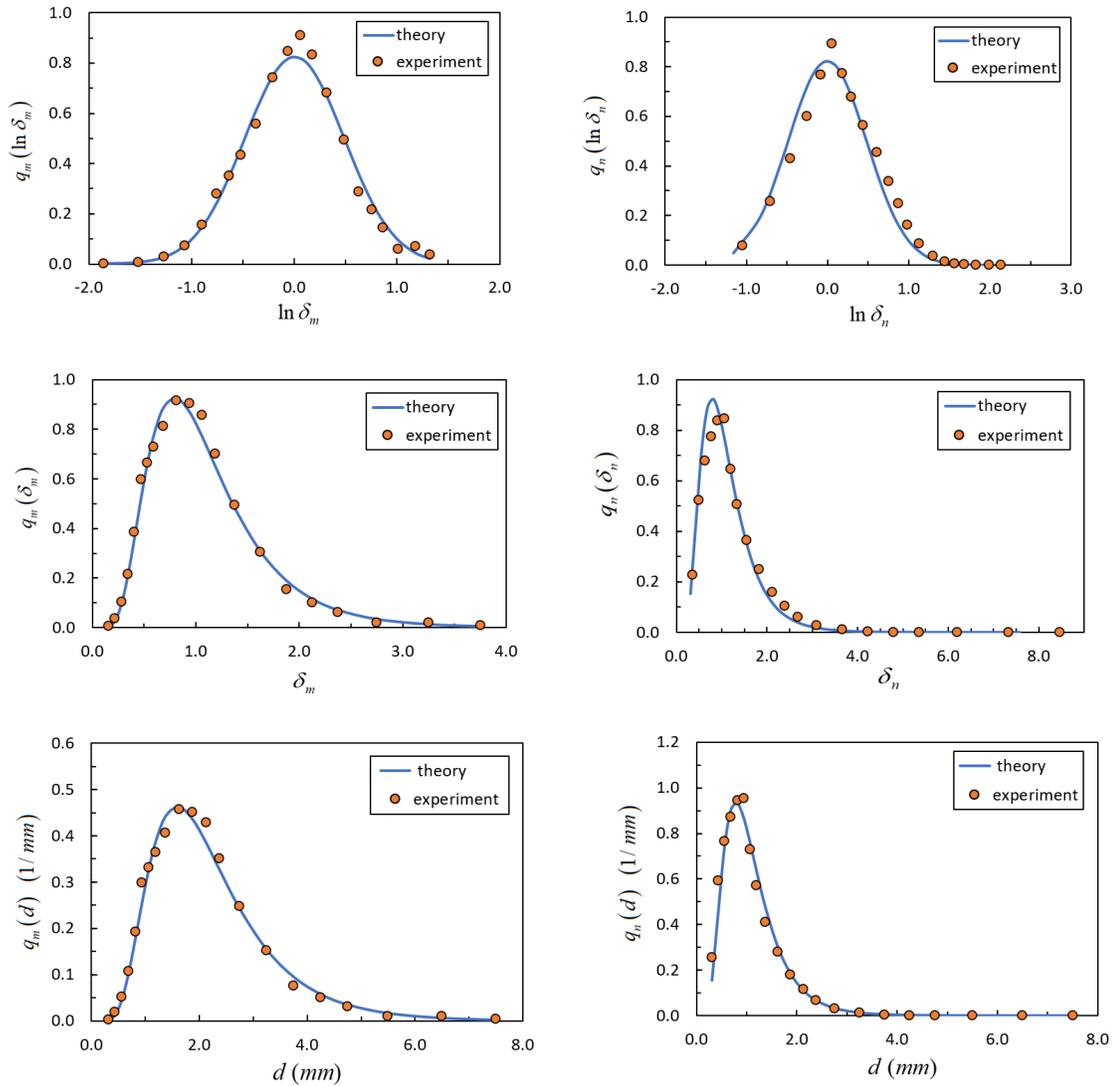

**FIG. S1.** Probability density functions of the drop-size distribution for convective rain with intensity $I = 4.5$ mm/h at China. Circles indicate experimental data from Niu et al. [3] and lines are theoretical calculations by Eqs. (S1)-(S9) with $\sigma_d = 0.485$.

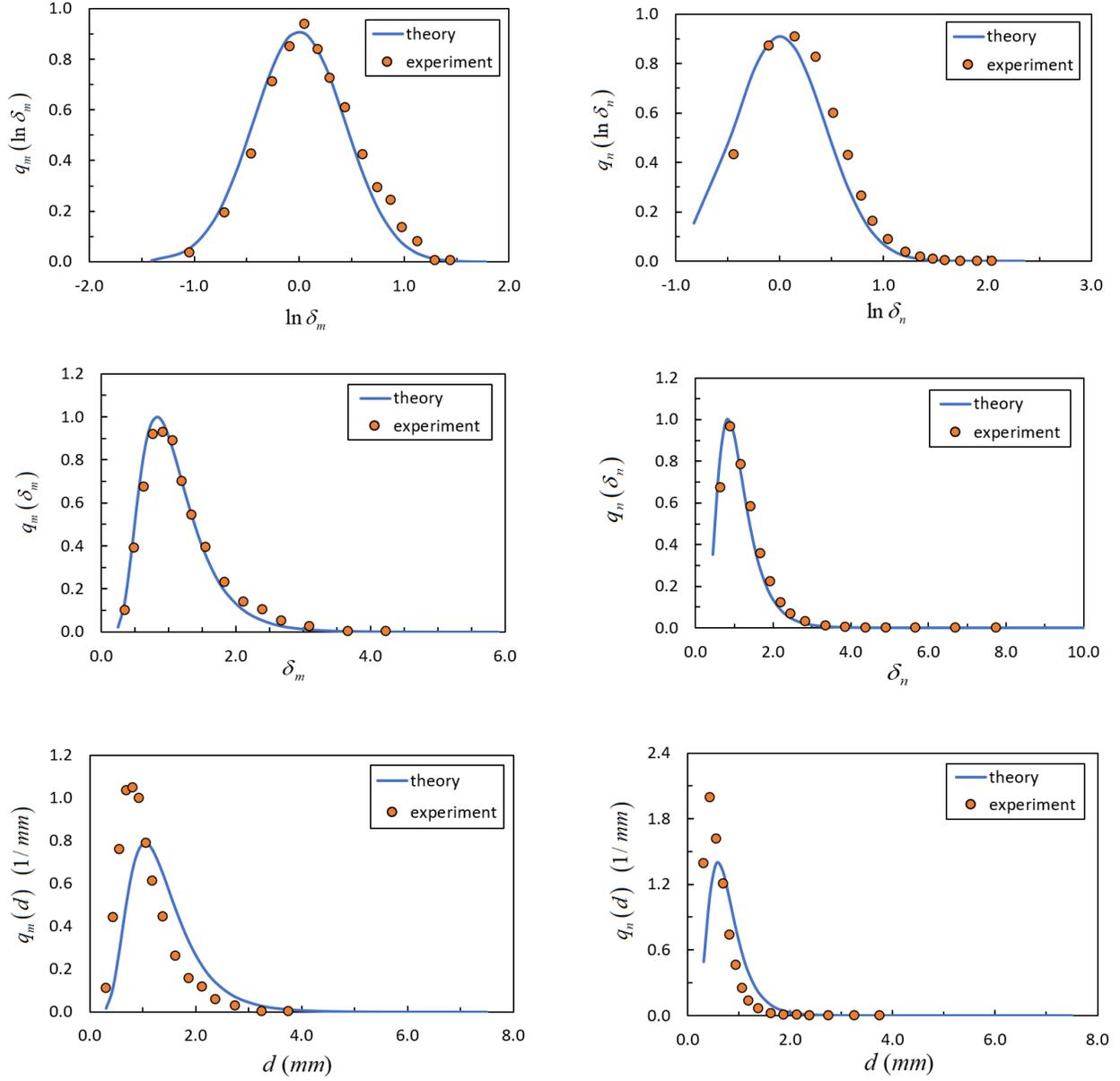

**FIG. S2.** Probability density functions of the drop-size distribution for stratiform rain with intensity $I$ = 0.5 mm/h at China. Circles indicate experimental data from Niu et al. [3] and lines are theoretical calculations by Eqs. (S1)-(S9) with $\sigma_d = 0.439$.

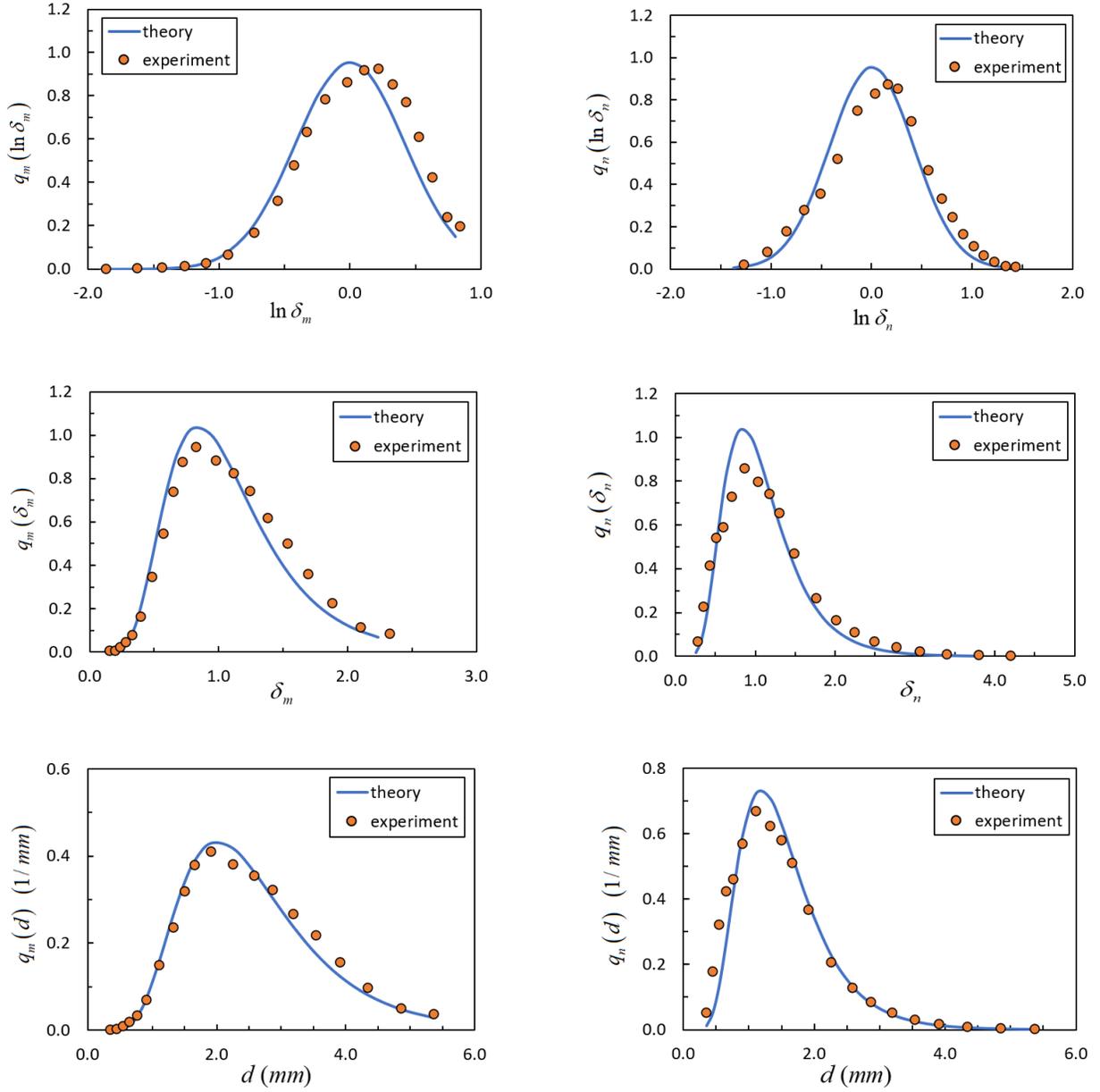

**FIG. S3.** Probability density functions of the drop-size distribution for convective rain with intensity $I$ = 9.3 mm/h at Greece. Circles indicate experimental data from Baltas et al. [4] and lines are theoretical calculations by Eqs. (S1)-(S9) with $\sigma_d = 0.419$.

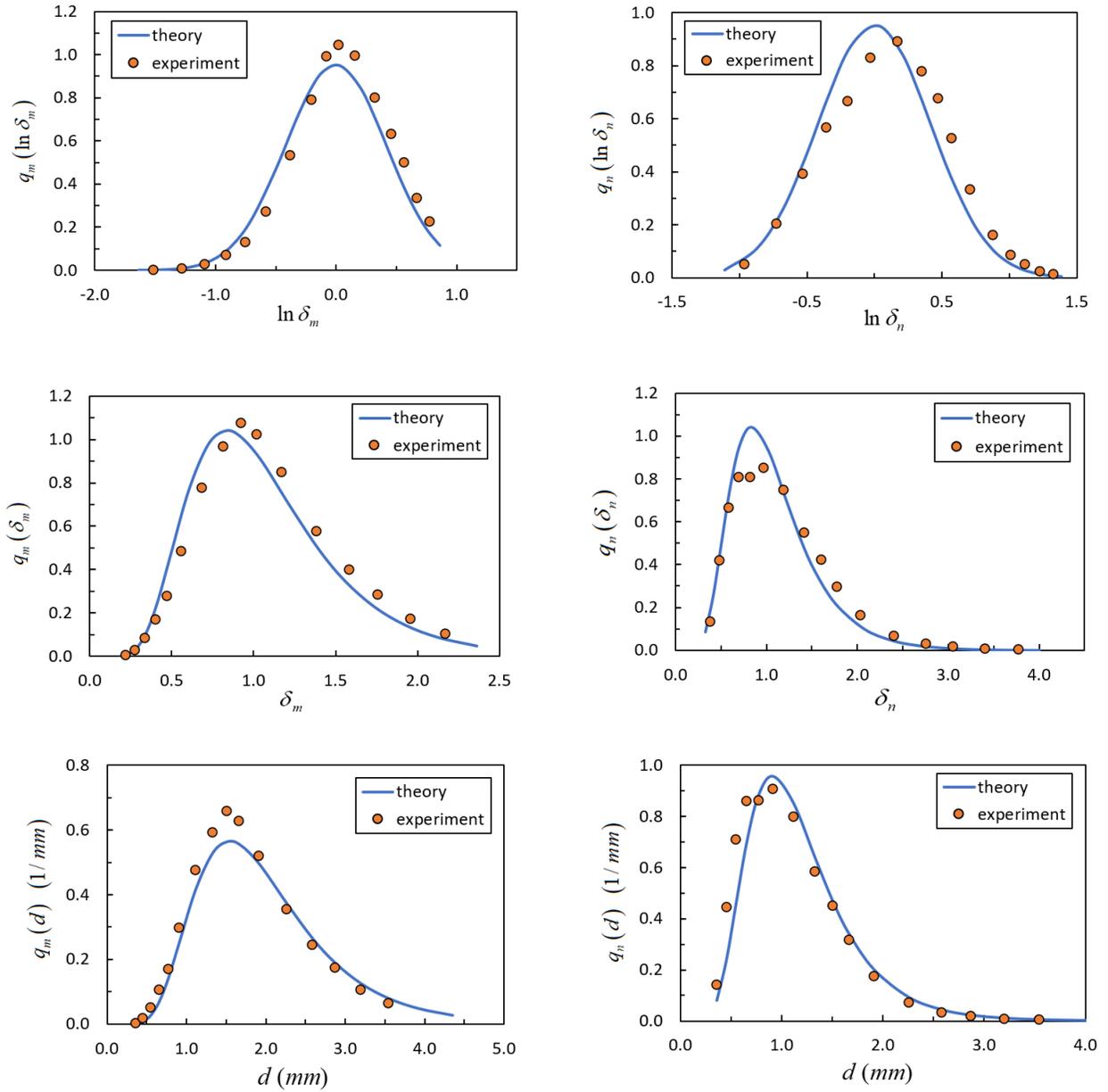

**FIG. S4.** Probability density functions of the drop-size distribution for stratiform rain with intensity $I$ = 2.8 mm/h at Greece. Circles indicate experimental data from Baltas et al. [4] and lines are theoretical calculations by Eqs. (S1)-(S9) with $\sigma_d = 0.419$.

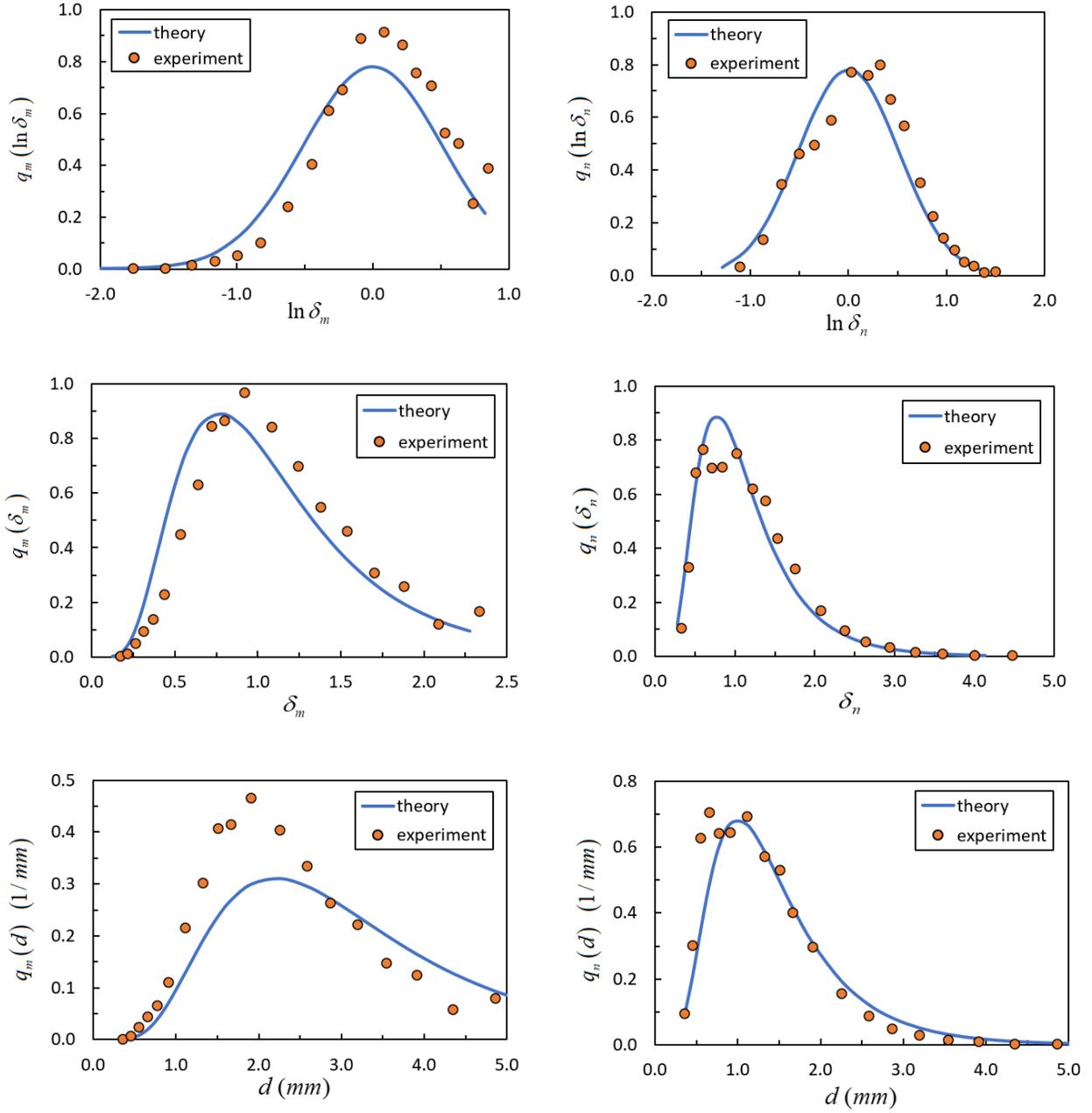

**FIG. S5.** Probability density functions of the drop-size distribution for rain with intensity $I$ = 22.9 mm/h at Malaysia. Circles indicate experimental data from Lam et al. [13] and lines are theoretical calculations by Eqs. (S1)-(S9) with $\sigma_d = 0.512$.

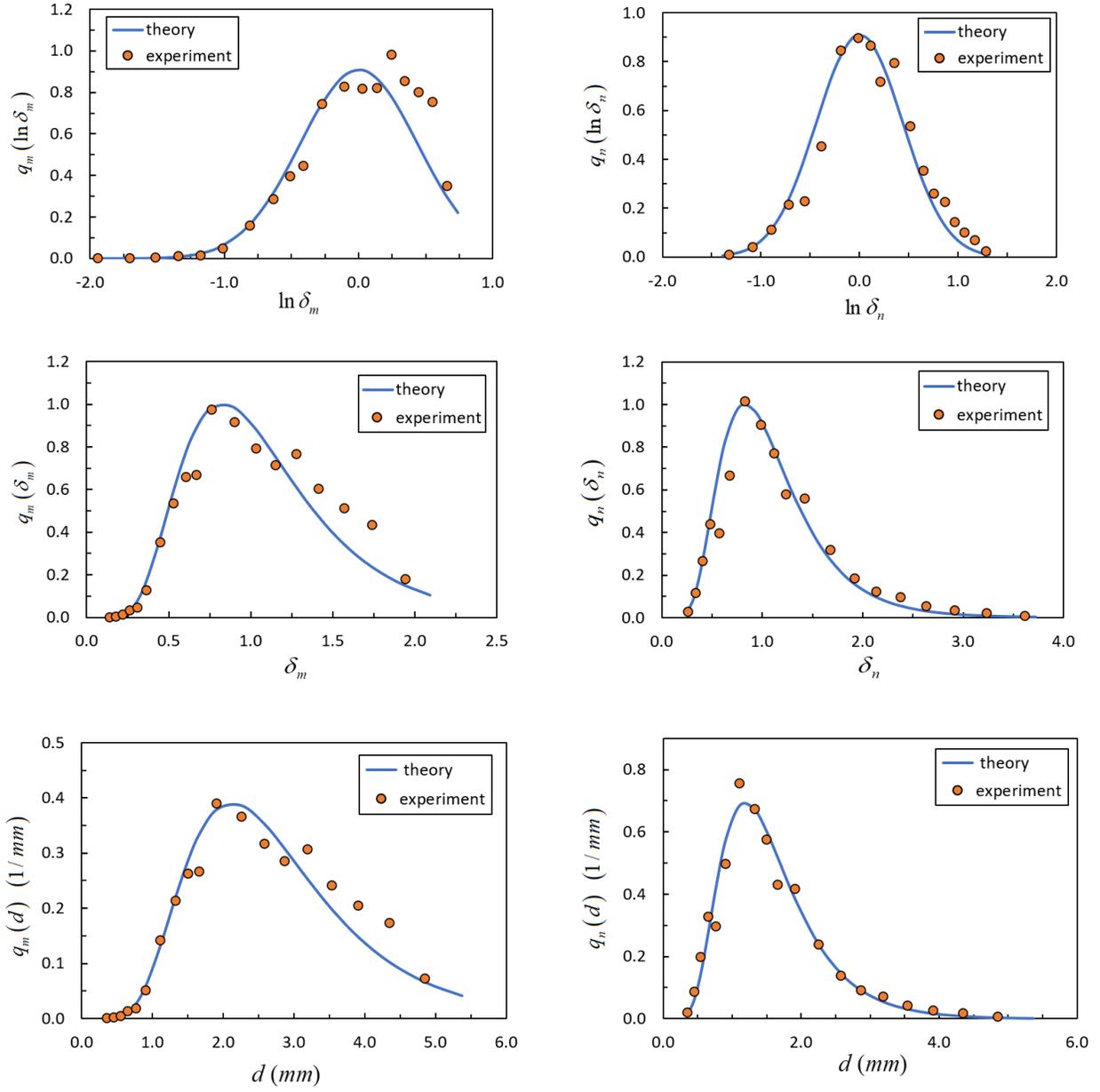

**FIG. S6.** Probability density functions of the drop-size distribution for convective rain with intensity $I$ = 12.9 mm/h at Amazon basin. Circles indicate experimental data from Tokay et al. [6] and lines are theoretical calculations by Eqs. (S1)-(S9) with $\sigma_d = 0.439$.

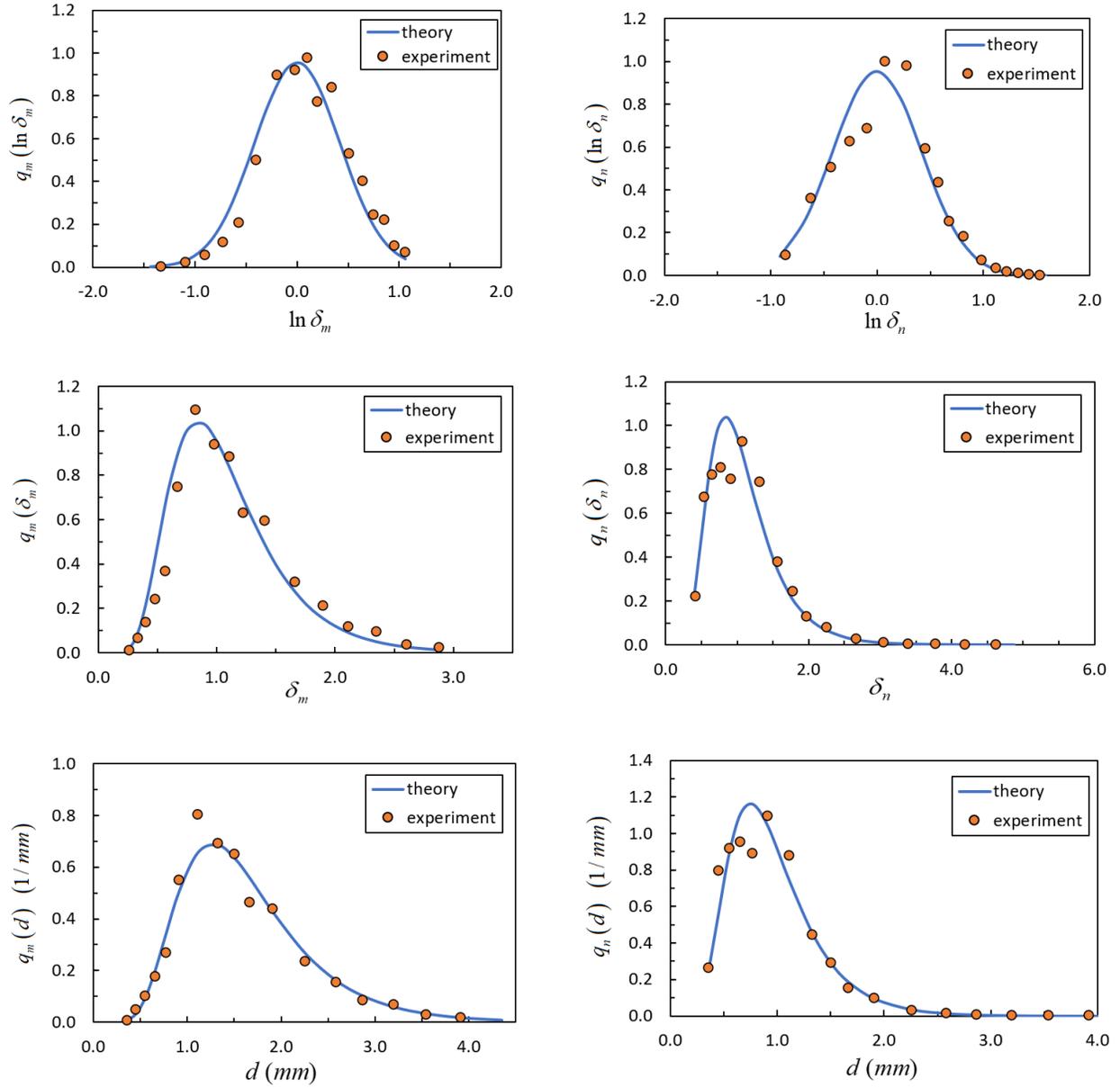

**FIG. S7.** Probability density functions of the drop-size distribution for stratiform rain with intensity $I$ = 1.15 mm/h at Amazon basin. Circles indicate experimental data from Tokay et al. [6] and lines are theoretical calculations by Eqs. (S1)-(S9) with $\sigma_d = 0.419$.

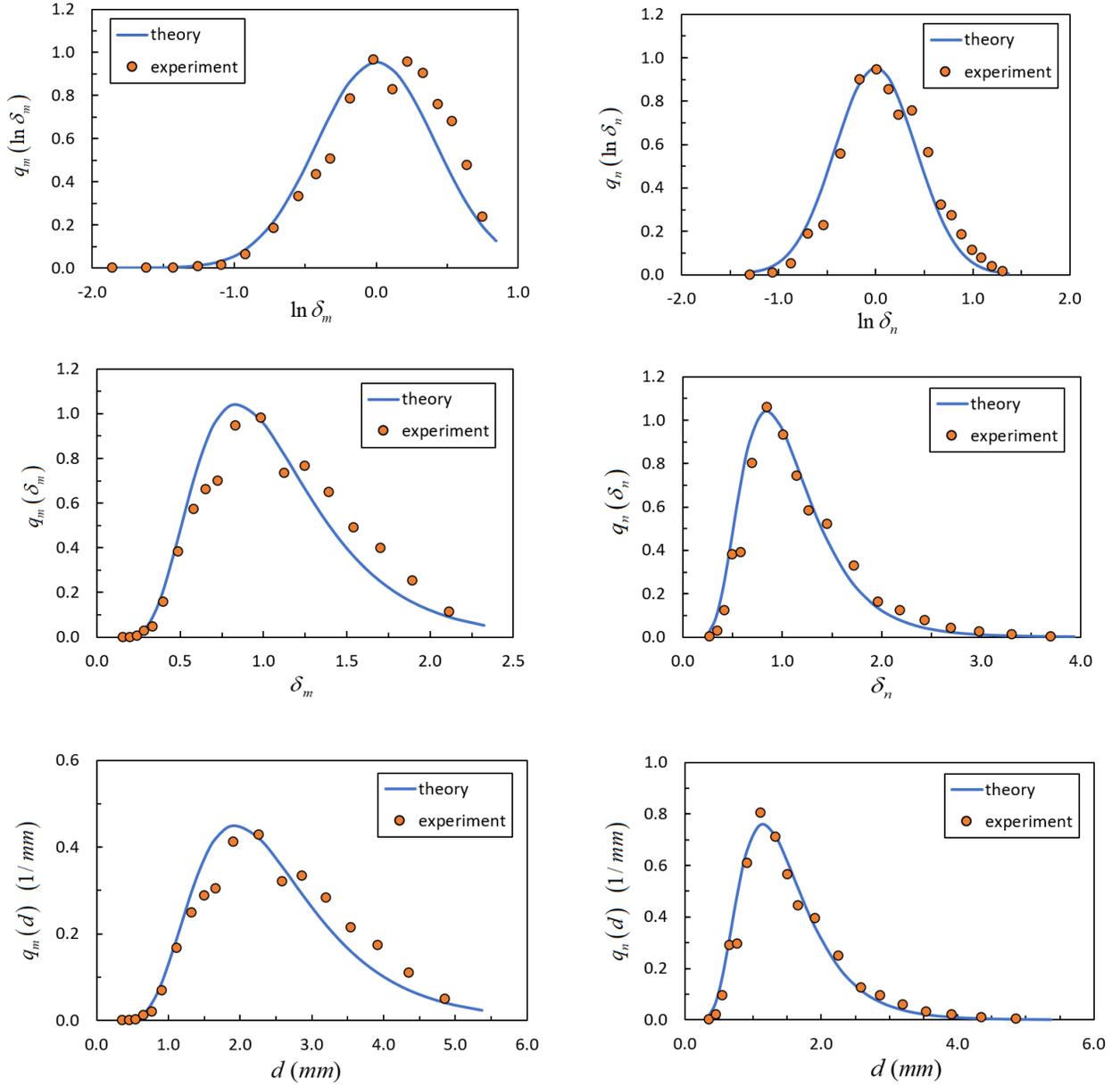

**FIG. S8.** Probability density functions of the drop-size distribution for convective rain with intensity $I$ = 7.8 mm/h at Florida, USA. Circles indicate experimental data from Tokay et al. [7] and lines are theoretical calculations by Eqs. (S1)-(S9) with $\sigma_d = 0.419$.

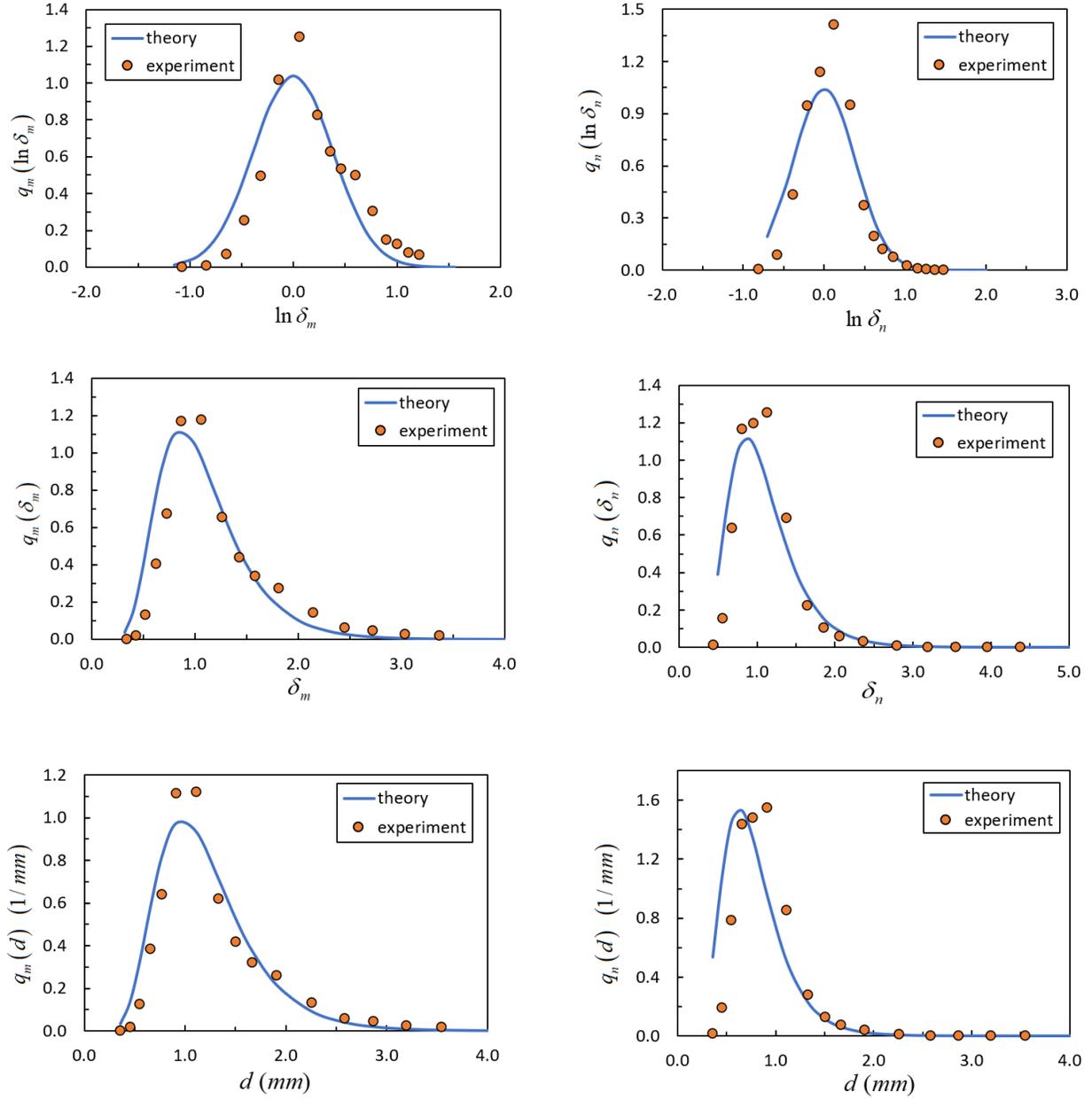

**FIG. S9.** Probability density functions of the drop-size distribution for convective rain with intensity $I$ = 0.3 mm/h at Florida, USA. Circles indicate experimental data from Tokay et al. [7] and lines are theoretical calculations by Eqs. (S1)-(S9) with $\sigma_d = 0.384$.

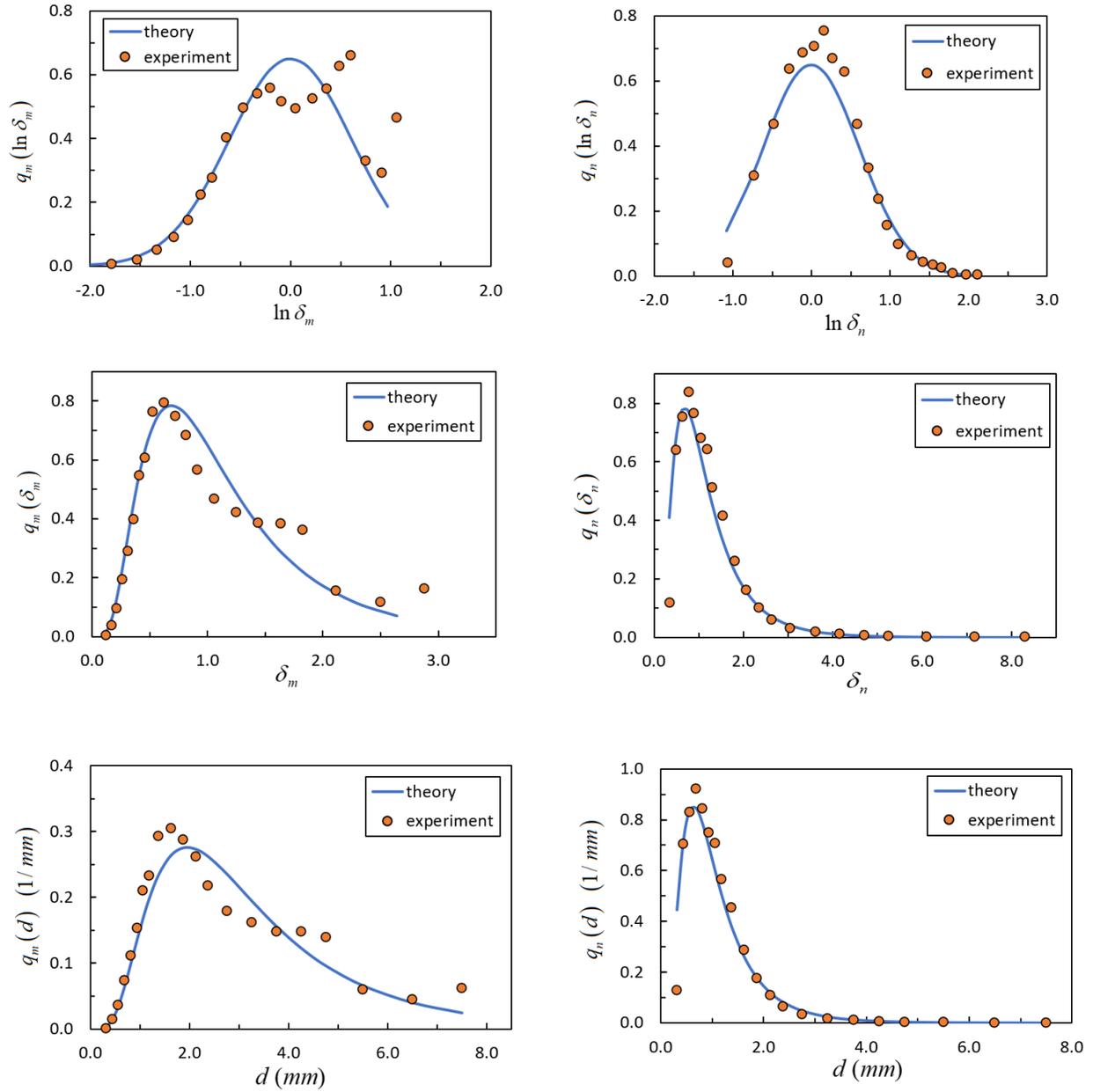

**FIG. S10.** Probability density functions of the drop-size distribution for convective rain with intensity $I$ = 26.76 mm/h at Jincheon, South Korea. Circles indicate experimental data from Loh et al. [8] and lines are theoretical calculations by Eqs. (S1)-(S9) with $\sigma_d = 0.614$.

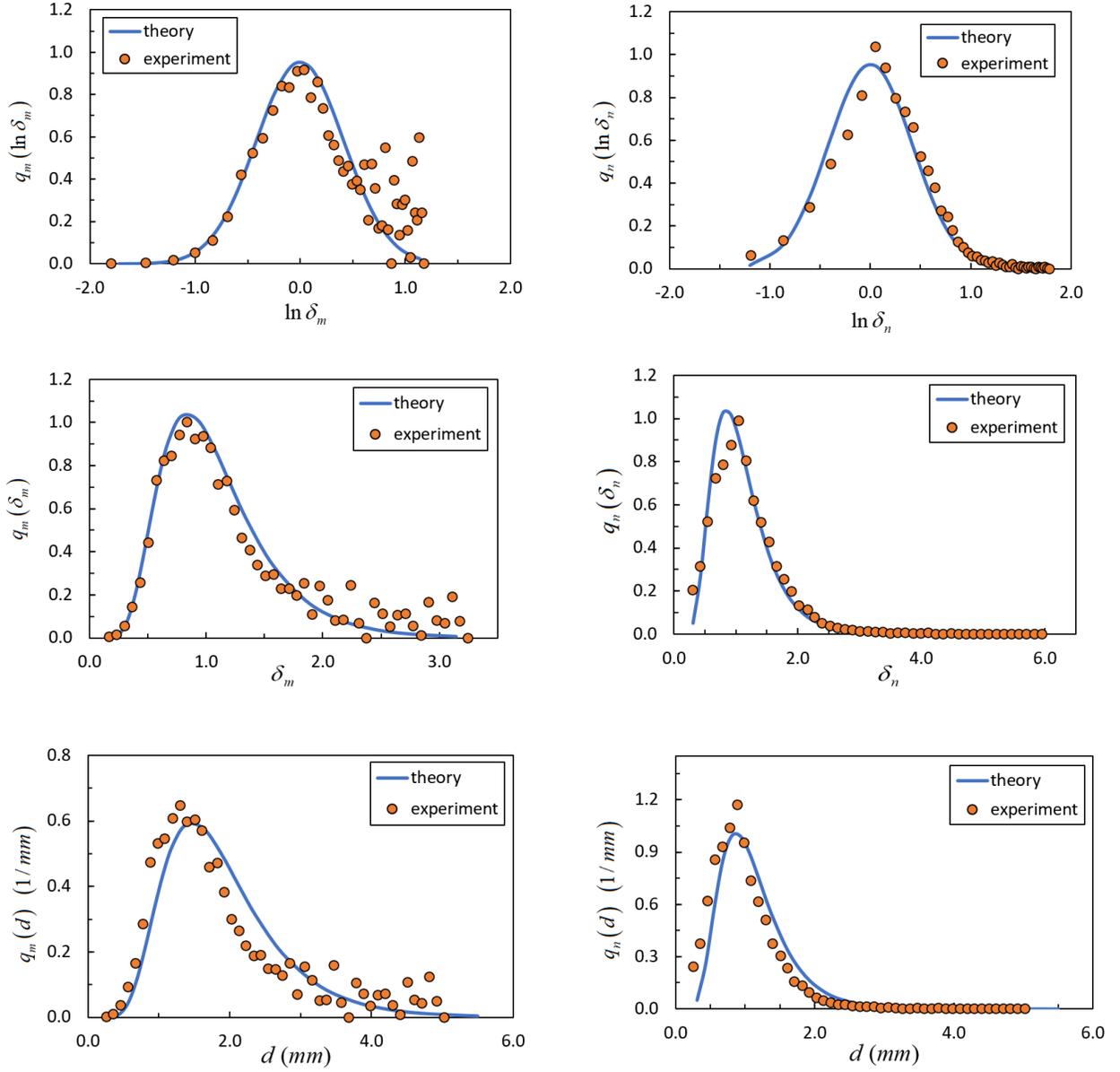

**FIG. S11.** Probability density functions of the drop-size distribution for convective rain with intensity $I$ = 2.24 mm/h at North Carolina, USA. Circles indicate experimental data from Smith [10] and lines are theoretical calculations by Eqs. (S1)-(S9) with $\sigma_d = 0.419$.

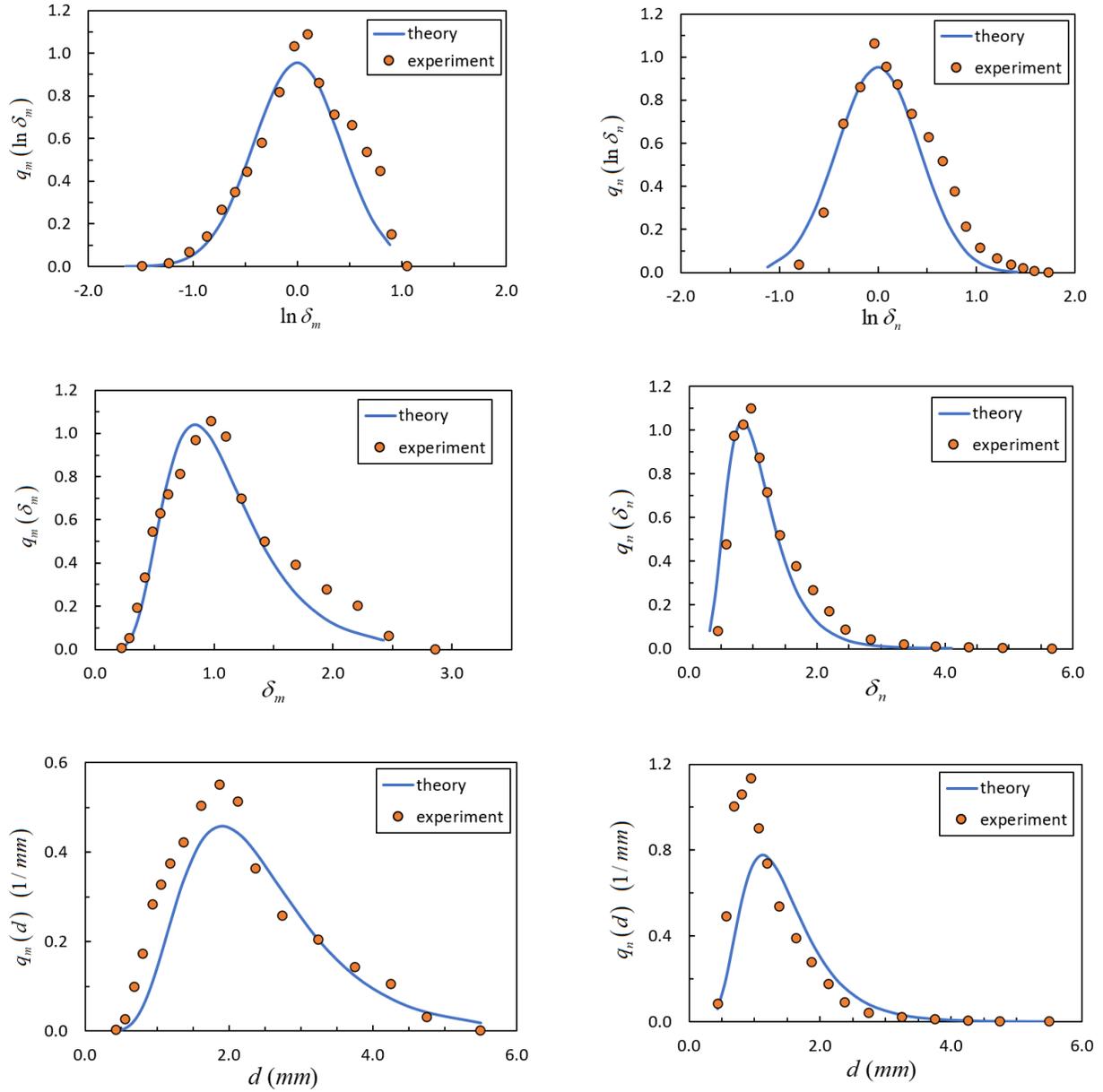

**FIG. S12.** Probability density functions of the drop-size distribution for rain with intensity $I$ = 7.2 mm/h at Alabama, USA. Circles indicate experimental data from Tokay et al. [5] and lines are theoretical calculations by Eqs. (S1)-(S9) with $\sigma_d = 0.419$.

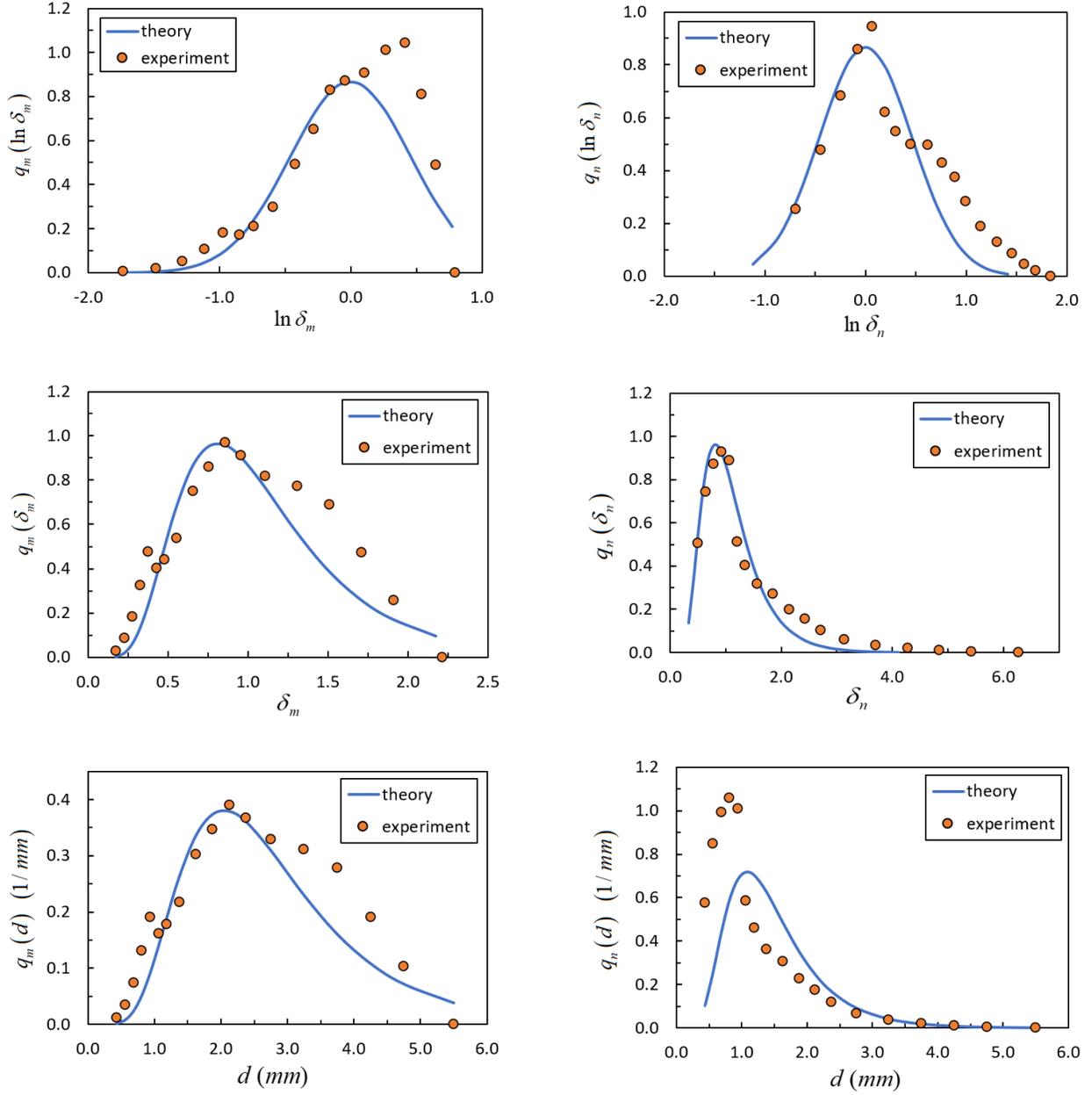

**FIG. S13.** Probability density functions of the drop-size distribution for rain with intensity $I$ = 12.4 mm/h at Alabama, USA. Circles indicate experimental data from Tokay et al. [5] and lines are theoretical calculations by Eqs. (S1)-(S9) with $\sigma_d = 0.461$.

## 2.3 Diagram of mass median diameters of raindrops

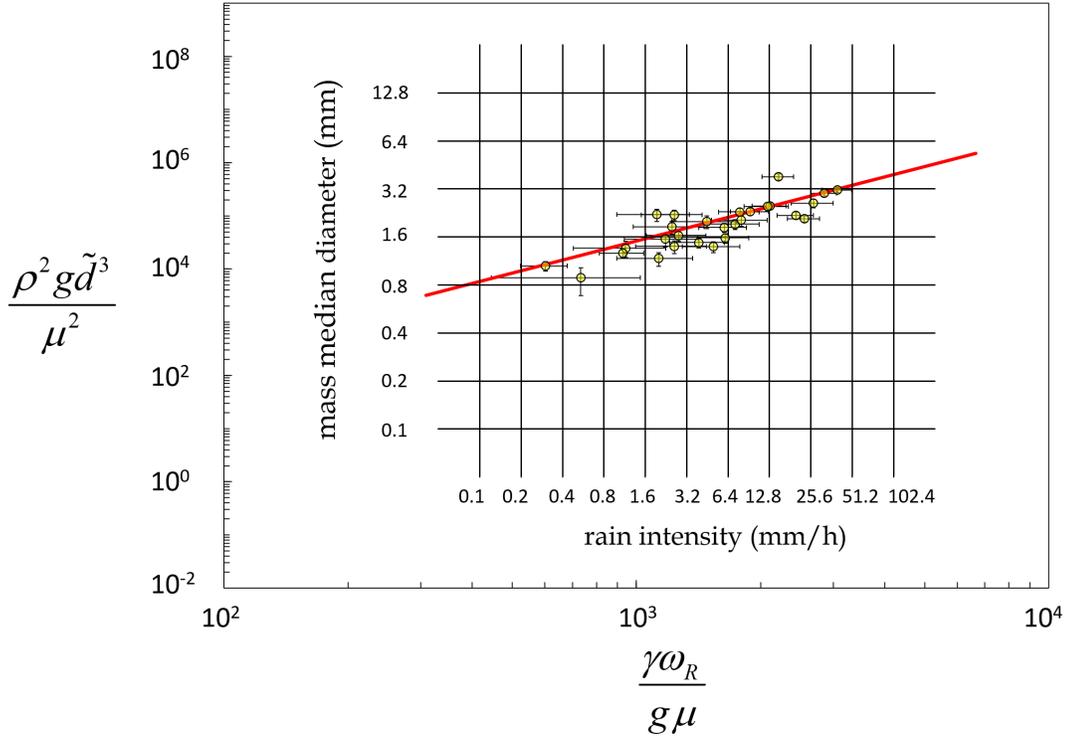

**FIG. S14.** Theoretical diagram of mass median diameters of raindrops as a function of rain intensity. The red solid line describes the theoretical mass median diameters, $\tilde{d}$, and is given by Eq. (7) of the main text. The circles correspond to the published experimental data, which were obtained from the analysis of time-averaged spectrograms of rains with intensities 0.4-40 mm/h recorded by disdrometers in different parts of the world (the respective reports are in the literature summarized in Table S1). For this purpose, the measured count median diameters, $\tilde{d}_n$, (see Table S1) were extracted from the rain spectrograms (see Appendix of the main text) and converted into mass median diameters using Eq. (S6). The error bars represent measurement uncertainties that we evaluated based on the disdrometer model and measurement error analysis [16,17]. The theoretical connection between the rain intensity and the rain frequency, $\omega_R$, is given by Eq. (9) of the main text, where the standard deviation, $\sigma_d$, of raindrop size distribution was assumed to be constant and was set to the geometric mean value $\sigma_d = \frac{1}{3\sqrt{2}} \ln 10$ for all rain intensities. The rainwater properties were set to $\rho = 1000$ kg/m$^3$, $\mu = 0.001$ Pa·s, $\gamma = 0.072$ N/m, and the gravitational acceleration $g = 9.81$ m/s$^2$.

**SI References**

[1] W. C. Hinds, *Aerosol Technology: Properties, Behavior, and Measurement of Airborne Particles* (Wiley, 2012).

[2] T. Hatch and S. P. Choate, *Statistical Description of the Size Properties of Non Uniform Particulate Substances*, J. Frankl. Inst. **207**, 369 (1929).

[3] S. Niu, X. Jia, J. Sang, X. Liu, C. Lu, and Y. Liu, *Distributions of Raindrop Sizes and Fall Velocities in a Semiarid Plateau Climate: Convective versus Stratiform Rains*, J. Appl. Meteorol. Climatol. **49**, 632 (2010).

[4] E. Baltas, D. Panagos, and M. Mimikou, *Statistical Analysis of the Raindrop Size Distribution Using Disdrometer Data*, Hydrology **3**, 1 (2016).

[5] A. Tokay, W. A. Petersen, P. Gatlin, and M. Wingo, *Comparison of Raindrop Size Distribution Measurements by Collocated Disdrometers*, J. Atmospheric Ocean. Technol. **30**, 1672 (2013).

[6] A. Tokay, A. Kruger, W. F. Krajewski, P. A. Kucera, and A. J. P. Filho, *Measurements of Drop Size Distribution in the Southwestern Amazon Basin*, J. Geophys. Res. Atmospheres **107**, LBA 19 (2002).

[7] A. Tokay, A. Kruger, and W. F. Krajewski, *Comparison of Drop Size Distribution Measurements by Impact and Optical Disdrometers*, J. Appl. Meteorol. Climatol. **40**, 2083 (2001).

[8] J. L. Loh, D.-I. Lee, M.-Y. Kang, and C.-H. You, *Classification of Rainfall Types Using PARSIVEL Disdrometer and S-Band Polarimetric Radar in Central Korea*, Remote Sens. **12**, 642 (2020).

[9] S.-H. Suh, H.-J. Kim, D.-I. Lee, and T.-H. Kim, *Geographical Characteristics of Raindrop Size Distribution in the Southern Parts of South Korea*, J. Appl. Meteorol. Climatol. **60**, 157 (2021).

[10] J. A. Smith, *Marked Point Process Models of Raindrop-Size Distributions*, J. Appl. Meteorol. Climatol. **32**, 284 (1993).

[11] L. P. D'Adderio, F. Porcù, and A. Tokay, *Evolution of Drop Size Distribution in Natural Rain*, Atmospheric Res. **200**, 70 (2018).

[12] Y. Zeng, L. Yang, Z. Tong, Y. Jiang, Z. Zhang, J. Zhang, Y. Zhou, J. Li, F. Liu, and J. Liu, *Statistical Characteristics of Raindrop Size Distribution during Rainy Seasons in Northwest China*, Adv. Meteorol. **2021**, 1 (2021).

[13] H. Y. Lam, J. Din, and S. L. Jong, *Statistical and Physical Descriptions of Raindrop Size Distributions in Equatorial Malaysia from Disdrometer Observations*, Adv. Meteorol. **2015**, e253730 (2015).

[14] J. W. Cha and S. S. Yum, *Characteristics of Precipitation Particles Measured by PARSIVEL Disdrometer at a Mountain and a Coastal Site in Korea*, Asia-Pac. J. Atmospheric Sci. **57**, 261 (2021).

[15] P. Fornasini, *The Uncertainty in Physical Measurements: An Introduction to Data Analysis in the Physics Laboratory* (Springer New York, 2008).

[16] J. L. Jaffrain and A. Berne, *Experimental Quantification of the Sampling Uncertainty Associated with Measurements from PARSIVEL Disdrometers*, J. Hydrometeorol. **12**, 19 (2011).

[17] J. Joss and A. Waldvogel, *Raindrop Size Distribution and Sampling Size Errors*, J. Atmospheric Sci. **26**, 566 (1969).

[18] R. Fraile, A. Castro, M. Fernández-Raga, C. Palencia, and A. I. Calvo, *Error in the Sampling Area of an Optical Disdrometer: Consequences in Computing Rain Variables*, https://doi.org/10.1155/2013/369450.

[19] H. S. Gertzman and D. Atlas, *Sampling Errors in the Measurement of Rain and Hail Parameters*, J. Geophys. Res. 1896-1977 **82**, 4955 (1977).

[20] A. Tokay, K. R. Wolff, P. Bashor, and O. K. Dursun, *On the Measurement Errors of the Joss-Waldvogel Disdrometer.*, in (NTRS - NASA Technical Reports Server, https://ntrs.nasa.gov/citations/20030053178, 2003), p. 4.

[21] A. Tokay, P. G. Bashor, and K. R. Wolff, *Error Characteristics of Rainfall Measurements by Collocated Joss–Waldvogel Disdrometers*, J. Atmospheric Ocean. Technol. **22**, 513 (2005).

[22] W.-Y. Chang, G. Lee, B. J.-D. Jou, W.-C. Lee, P.-L. Lin, and C.-K. Yu, *Uncertainty in Measured Raindrop Size Distributions from Four Types of Collocated Instruments*, Remote Sens. **12**, 1167 (2020).

[23] S. L. M. Neto, Y. Sakagami, I. C. da Costa, E. B. Pereira, J. C. Thomaz Junior, and A. Von Wangenheim, Uncertainty Analysis of Disdrometer Model PARSIVEL2 for Rainfall Amount, Joint Research Report of INPECCST, INPE-CPTEC, UFSCEMC- LEPTEN/LABSOLAR, IFSC-Florianópolis and UFSCINE- LAPIX/INCOD No. o sid.inpe.br/mtc-m21b/2016/02.12.11.38-RPQ, INPE, 2015.